\documentclass[twocolumn]{aastex631} % linenumbers

\usepackage{amsmath}
\usepackage{rotating}
\usepackage{float}

\begin{document}

\title{Meteoroid stream identification with HDBSCAN unsupervised clustering algorithm}

\correspondingauthor{Eloy Peña-Asensio}
\email{eloy.pena@polimi.it}

\author[0000-0002-7257-2150]{Eloy Peña-Asensio}
\affiliation{Department of Aerospace Science and Technology, Politecnico di Milano, Via La Masa 34, 20156 Milano, Italy}
%\email[show]{eloy.pena@polimi.it}  

\author[0000-0001-7537-4996]{Fabio Ferrari} 
\affiliation{Department of Aerospace Science and Technology, Politecnico di Milano, Via La Masa 34, 20156 Milano, Italy}
%\email{fabio1.ferrari@polimi.it}



\begin{abstract}
Accurate identification of meteoroid streams is central to understanding their origins and evolution. However, overlapping clusters and background noise hinder classification, an issue amplified for missions such as ESA’s LUMIO that rely on meteor shower observations to infer lunar meteoroid impact parameters. This study evaluates the performance of the Hierarchical Density-Based Spatial Clustering of Applications with Noise (HDBSCAN) algorithm for unsupervised meteoroid stream identification, comparing its outcomes with the established Cameras for All-Sky Meteor Surveillance (CAMS) look-up table method. We analyze the CAMS Meteoroid Orbit Database v3.0 using three feature vectors: LUTAB (CAMS geocentric parameters), ORBIT (heliocentric orbital elements), and GEO (adapted geocentric parameters). HDBSCAN is applied with varying minimum cluster sizes and two cluster selection methods (\textit{eom} and \textit{leaf}). To align HDBSCAN clusters with CAMS classifications, the Hungarian algorithm determines the optimal mapping. Clustering performance is assessed via the Silhouette score, Normalized Mutual Information, and F1 score, with Principal Component Analysis further supporting the analysis. With the GEO vector, HDBSCAN confirms 39 meteoroid streams, 21 strongly aligning with CAMS. The ORBIT vector identifies 30 streams, 13 with high matching scores. Less active showers pose identification challenges. The \textit{eom} method consistently yields superior performance and agreement with CAMS. Although HDBSCAN requires careful selection of the minimum cluster size, it delivers robust, internally consistent clusters and outperforms the look-up table method in statistical coherence. These results underscore HDBSCAN’s potential as a mathematically consistent alternative for meteoroid stream identification, although further validation is needed to assess physical validity.

\end{abstract}


%-------------------------------------------------------------------
\section{Introduction}

Meteoroids are small solid particles (from 30 micrometers to 1 meter in size) derived from comets, asteroids, or other planetary objects, orbiting the Sun \citep{Koschny2017JIMO4591K}. Upon entering Earth's atmosphere, these particles undergo rapid heating and ablation, producing a luminous phenomenon known as a meteor or fireball, depending on their brightness \citep{Ceplecha1998SSRv, Silber2018AdSpR}. Meteoroid streams are groups of meteoroids originated from a common progenitor that share similar orbital parameters. These streams are the result of ejection mechanisms, including sublimation or fragmentation, leading to dispersal over time and ultimately contributing to the sporadic meteoroid background \citep{Murray1982Icar49125M, Babadzhanov1987PAICz67141B, Jenniskens1998EPS50555J, Pauls2005MandPS40241P, Vaubaillon2019msmebook161V}. When Earth crosses the path of a meteoroid stream, a meteor shower occurs, characterized by increased meteor activity radiating from a specific point in the sky, the radiant, and similar velocities \citep{Jenniskens2006mspcbookJ, Jenniskens2023Atlas}. The dynamics of meteoroid streams, mainly originated by comet disruptions \citep{Jenniskens2008Icar19413J, Jenniskens2008book, Nesvorny2010ApJ713816N, YangIshiguro2015ApJ81387Y, Durisova2024MNRAS5353661D}, are influenced by gravitational perturbations, solar radiation pressure, and inter-particle collisions, producing complex and evolving distributions \citep{Williams2014me13conf179W}.

The accurate association of individual meteors with their parent bodies or streams remains challenging due to (a) overlapping orbital parameters, (b) the presence of sporadic meteor activity, (c) observational constraints, and (d) the limitations of the methods employed, making it an ongoing area of research \citep{Nesluvan2017AA598A40N, Jopek2024AA682A159J, Eloy2024AdSpR741073P, Shober2024AA686A130S, Shober2025AA693A23S}. The Meteor Data Center (MDC) of the International Astronomical Union (IAU) compiles published information on meteor showers \citep{Hajdukova2023AA671A155H}. The MDC curates two lists of average meteoroid stream parameters, including orbital elements, shower activity, and radiant properties: one of confirmed (established) showers and a working list of proposed showers. Unfortunately, many parameters of the showers remain poorly constrained, as radiant positions and orbital elements reported by different authors often show significant discrepancies, mainly because of observational uncertainties and limited data samples and observational limitations. 

Meteoroids streams and meteor showers are interconnected phenomena that encapsulate the interaction of interplanetary matter with the Sun, planets, and the Earth's atmosphere \citep{Koschny2019SSRv21534K}. Understanding their formation, composition, and identification is central to meteor science, with implications ranging from planetary defense to the study of the Solar System and beyond \citep{Turchak2014JTAM44d15T, Moreno2015Icar250544M, Eloy2021MNRAS5044829P, Eloy2022AJ16476P, Eloy2023MNRAS5205173P, DenisVida2023NatAs7318V, Eloy2024Icar40815844P, Eloy2024PSJ5206P, Hemmelgarn_2024}. Proper identification of meteoroid streams is needed for determining the mean physical properties of their members, which is particularly important when assumptions about the impactor are required, as in Lunar Impact Flash (LIF) analysis \citep{BellotRubio1998EMP82575B, Suggs2008EMP102293S, Bouley2012Icar218115B, Ortiz2015MNRAS454344O, Madiedo2015PSS111105M, Bonanos2018AA612A76B, Avdellidou2019MNRAS4845212A, Liakos2024AA687A14L}. In this regard, accurate characterization of meteoroid streams is critical for the European Space Agency’s (ESA) Lunar Meteoroid Impact Observer (LUMIO) mission, which will monitor and study LIFs on the Moon's farside \citep{Cipriano2018FrASS529C, Topputo2023Icar38915213T}.

To this end, similarity orbital criteria have been developed to systematically identify meteoroid streams by quantifying the resemblance between the orbital parameters of individual meteoroids. One of the first and most widely utilized criteria is the Southworth-Hawkins D-function ($D_{SH}$), introduced to assess the degree of orbital similarity between meteoroids or between a meteoroid and a parent body \citep{Southworth1963SCoA7261S}. The $D_{SH}$ parameter is calculated using five orbital elements, including the perihelion distance ($q$), eccentricity ($e$), inclination ($i$), argument of perihelion ($\omega$), and longitude of ascending node ($\Omega$): 

%\begin{equation}
%D_{SH}^{2}=\left(e_{B}-e_{A}\right)^{2}+\left(q_{B}-q_{A}\right)^{2}+\left(2 \sin \frac{I_{AB}}{2}\right)^{2}
%+\left(\frac{e_{B}+e_{A}}{2}\right)^{2}\left(2 \sin \frac{\pi_{B A}}{2}\right)^{2},
%	\label{eq:D_SH}
%\end{equation}

\begin{align}
D_{SH}^{2} &= (e_{B}-e_{A})^{2} + (q_{B}-q_{A})^{2}
     + \bigl(2\sin\tfrac{I_{AB}}{2}\bigr)^{2} + \notag\\
&\quad + 
     \Bigl(\tfrac{e_{B}+e_{A}}{2}\Bigr)^{2}
     \bigl(2\sin\tfrac{\pi_{BA}}{2}\bigr)^{2}\,,
\end{align}

where additional geometric concepts are considered, including $\pi_{BA}$, the difference between the longitudes of perihelion measured from the intersection of the orbits, and the angle representing the difference in their orbital inclinations ($I_{AB}$). Subsequent efforts by \citet{Drummond1981Icar45545D} expanded upon this framework, introducing refinements and defining the widely used $D_D$. However, these D-functions do not satisfy the mathematical definition of metrics because they fail to fulfill the triangle inequality \citep{Kholshevnikov2016MNRAS}. Instead, they are better described as quasimetrics, as they comply with a generalized or relaxed version of the triangle inequality \citep{Milanov2019CeMDA}. This approach necessitates careful inspection of the threshold value to establish accurate boundaries for association. Near-ecliptic orbits, characterized by higher meteoroid densities and reduced differentiation in inclination, require lower thresholds to minimize contamination by sporadic meteors \citep{Galligan2001MNRAS327623G}. Moreover, \citet{Shober2025AA693A23S} investigated the dependence on D-function thresholds and found that orbital similarities between fireballs and parent bodies or streams were consistently indistinguishable from random associations due to the inherent chaotic dynamics and rapid decoherence of near-Earth orbits.

In contrast to the computational complexity and limitations of D-functions for associating meteors with parent showers, a look-up table approach provides an intuitive alternative \citep{Jenniskens2018PSS15421J}. This method relies on a precompiled table of meteor characteristics. The table includes solar longitude, Sun-centered ecliptic radiant coordinates, geocentric velocity ($V_{\mathrm{g}}$), and the IAU shower number. By condensing these key parameters, the look-up table facilitates rapid shower identification based on observable meteor properties. In addition to the look-up table method, novel approaches have emerged that exploit machine learning. \citet{Eloy2024AdSpR741073P} demonstrated that the efficacy of machine learning distance metrics, such as standardized Euclidean and Bray-Curtis, is similar or even outperform traditional D-functions in distinguishing meteors from the sporadic background. Another example is \citet{Sugar2017MPS}, who applied the Density-Based Spatial Clustering of Applications with Noise (DBSCAN) algorithm \citep{1996kddmconf226E} for unsupervised identification of meteor showers. A more recent application of DBSCAN to identify potential meteor clusters is presented in \citet{Ashimbekova2025arXiv250316157A}.

In this study, we build upon the foundational work of \citet{1996kddmconf226E}; however, while DBSCAN has proven effective for density-based clustering, it is limited by its reliance on fixed density parameters, reducing its adaptability to datasets with varying density distributions. To address this limitation, we extend the analysis by evaluating the performance of its successor, Hierarchical Density-Based Spatial Clustering of Applications with Noise (HDBSCAN) \citep{Campello2013}. HDBSCAN extends DBSCAN by eliminating the need for fixed density thresholds and constructing a hierarchical clustering model that identifies clusters across multiple density levels. Although it still requires one parameter ---the minimum cluster size---, it removes the subjective considerations inherent in traditional for meteor shower classification. In this study, we assess HDBSCAN's capability for unsupervised meteoroid stream identification.

%-------------------------------------------------------------------
\section{Methods} \label{sec:methods}

This section describes the methodology used to analyze and evaluate the clustering performance of HDBSCAN in the context of meteoroid stream identification. First, we introduce the dataset utilized in this study, based on the publicly available CAMS meteoroid orbit catalog, along with the filtering criteria applied to ensure data reliability (\autoref{sec:data}). We then define the input feature vectors used for clustering, including the CAMS look-up table vector (\autoref{sec:lutab}), the orbital parameter-based vector (\autoref{sec:orbit}), and the geocentric vector representation (\autoref{sec:geo}). To assess the internal structure and variance of the dataset, we apply Principal Component Analysis (PCA) before performing clustering (\autoref{sec:pca}). 

HDBSCAN is employed as the clustering algorithm, with its framework and parameter selection strategy outlined in \autoref{sec:hdbscan}. The algorithm’s efficiency is enhanced by leveraging Borůvka’s algorithm for Minimum Spanning Tree (MST) construction (\autoref{sec:boruvka}) and the Ball Tree data structure for fast nearest-neighbor searches (\autoref{sec:balltree}). 

We do not validate HDBSCAN against a synthetic meteoroid population in this study because DBSCAN has already been benchmarked in a highly similar context by \citet{Sugar2017MPS}. Extensive literature has consistently demonstrated the superior performance of HDBSCAN over DBSCAN in terms of robustness, noise handling, and cluster stability \citep{Campello2013, Campello2015, McInnes2017, Ibrahim2018, Zhang2018, Ghamarian2019, Lentzakis2020, Tran2021, Wang2021, Mishra2022, Balducci2024, Wang2025}.

To evaluate clustering performance, we compute multiple metrics. The Silhouette score is used to assess the compactness and separation of clusters (\autoref{sec:silhouette}), while the Normalized Mutual Information score (NMI) quantifies the agreement between HDBSCAN-generated clusters and the CAMS reference classifications (\autoref{sec:nmi}). To establish an optimal mapping between the detected clusters and known meteor showers, we implement the Hungarian algorithm (\autoref{sec:hungarian}). Finally, the F1 score is calculated to measure the accuracy of these assignments, providing a comprehensive assessment of clustering quality (\autoref{sec:f1_score}).

\subsection{Baseline data} \label{sec:data}

CAMS (Cameras for All-Sky Meteor Surveillance) is an international project managed by the Carl Sagan Center at the SETI Institute, USA, with sponsorship from NASA \citep{Jenniskens2011Icar}. Its primary objective is to systematically monitor and map meteor activity through nighttime optical video surveillance combined with triangulation techniques. The last publicly released dataset, the Meteoroid Orbit Database v3.0, contains 471,582 meteor orbits recorded between 2010 and 2016 \citep{Jenniskens2018PSS15421J}, which forms the basis of the present work. 

While other automated meteor detection networks exist, CAMS remains one of the most widely recognized and extensively utilized repositories for meteor orbit data. However, its performance in detecting fast meteors has been found to be less reliable compared to its accuracy for slower meteors \citep{Koseki2017eMetN, Koseki2022eMetN}. To address these limitations and improve dataset quality, we implemented a filtering process to exclude lower-quality detections and minimize spurious results. The filtering criteria included: (1) a minimum convergence angle of 15$^\circ$ between camera detections to ensure reliable triangulation, (2) a velocity error not exceeding 10\% of the nominal value, (3) restriction to non-hyperbolic orbits ($e \leq 1$), and (4) a perihelion distance compatible with Earth-impacting orbits ($q \leq 1$ au). To ensure statistical significance, we exclude meteor showers with fewer than 100 members, resulting in a dataset of 316,235 meteors distributed across 76 unique meteor showers, with approximately 70\% classified as sporadic events. %We consider the classifications provided by CAMS to be accurate, which may not be necessarily true.

\subsection{Input vectors for clustering} \label{sec:vectors}

To evaluate the performance of HDBSCAN in identifying meteor showers, we utilize three distinct vector representations: LUTAB, ORBIT, and GEO.

To account for the periodic nature of angular elements and avoid discontinuities at the $0^\circ/360^\circ$ boundary, we represent the angular parameters using their sine and cosine components. The factor $1/\sqrt{2}$ is applied to the sine and cosine components to ensure that the combined contribution of each angular element remains comparable to that of the scalar components. This avoids overweighting angular elements in the Euclidean distance metric used for clustering.

After constructing the vector, each component is standardized to zero mean and unit variance across the dataset using Z-score normalization. This process ensures that all features contribute equally to the clustering process, regardless of their original units or dynamic ranges. For each component $x_i$, the standardized value $x_i'$ is computed as:

\begin{equation}
    x_i' = \frac{x_i - \mu_i}{\sigma_i},
\end{equation}

where $\mu_i$ and $\sigma_i$ are the mean and standard deviation of feature $x_i$ over the entire dataset. Note that Z-score standardization equalizes numerical scale without altering the covariance structure of the parameters; hence no assumption of statistical independence is implied, and physical correlations between features are preserved.

\subsubsection{CAMS look-up table vector} \label{sec:lutab}

In the CAMS database, shower assignment was performed manually by comparing newly observed meteors to a look-up table based on four parameters, which we refer to as the LUTAB vector:

\begin{equation}
\label{eqn:LUTAB}
LUTAB =
\left[\begin{array}{c}
\frac{1}{\sqrt{2}}\sin\lambda_{\odot} \\
\frac{1}{\sqrt{2}}\cos\lambda_{\odot} \\
\frac{1}{\sqrt{2}}\sin\alpha_{\odot} \\
\frac{1}{\sqrt{2}}\cos\alpha_{\odot} \\
\beta_{\odot} \\
V_{\mathrm{g}}
\end{array}\right]
=
\left[\begin{array}{c}
\frac{1}{\sqrt{2}}\sin\lambda_{\odot} \\
\frac{1}{\sqrt{2}}\cos\lambda_{\odot} \\
\frac{1}{\sqrt{2}}\sin(\lambda_{\mathrm{g}} - \lambda_{\odot}) \\
\frac{1}{\sqrt{2}}\cos(\lambda_{\mathrm{g}} - \lambda_{\odot}) \\
\beta_{\mathrm{g}} \\
V_{\mathrm{g}}
\end{array}\right],
\end{equation}

where \(V_{\mathrm{g}}\) represents the geocentric velocity of the meteoroid, $\lambda_{\odot}$ is the solar ecliptic longitude, and the pair $(\lambda_{\mathrm{g}} - \lambda_{\odot},\, \beta_{\mathrm{g}})$ represents the ecliptic longitude and latitude of the radiant in a Sun-centered reference frame. In some sources, the Sun-centered ecliptic longitude is denoted as $\alpha_{\odot} = \lambda_{\mathrm{g}} - \lambda_{\odot}$. The association with meteor showers was determined by CAMS using straightforward criteria: the observed solar longitude must be within $\pm$5$^\circ$, the Sun-centered radiant coordinates within $\pm$1$^\circ$, and geocentric velocity within $\pm$1 km/s.

\subsubsection{D-functions vector} \label{sec:orbit}

To evaluate the performance of HDBSCAN, we define the ORBIT vector based on the five heliocentric orbital elements used in traditional D-functions allowing a straightforward comparison. 

This results in a 7-dimensional feature vector:
\begin{equation}
\label{eqn:ORBIT}
ORBIT =
\left[\begin{array}{c}
    q \\
    e \\
    i \\
    \frac{1}{\sqrt{2}}\sin(\omega) \\
    \frac{1}{\sqrt{2}}\cos(\omega) \\
    \frac{1}{\sqrt{2}}\sin(\Omega) \\
    \frac{1}{\sqrt{2}}\cos(\Omega)
\end{array}\right].
\end{equation}

\subsubsection{Geocentric vector} \label{sec:geo}

The third vector used to evaluate HDBSCAN performance is the GEO vector, as proposed by \citet{Sugar2017MPS}. The GEO vector, adapted from geocentric parameters, represents a six-component vector:

\begin{equation}
\label{eqn:GEO}
GEO=
\left[\begin{array}{c}
\cos \left(\lambda_{\odot}\right) \\
\sin \left(\lambda_{\odot}\right) \\
\sin \left(\lambda_{\mathrm{g}}-\lambda_{\odot}\right) \cos \left(\beta_{\mathrm{g}}\right) \\
\cos \left(\lambda_{\mathrm{g}}-\lambda_{\odot}\right) \cos \left(\beta_{\mathrm{g}}\right) \\
\sin \left(\beta_{\mathrm{g}}\right) \\
V_{\mathrm{g}} / 72
\end{array}\right].
\end{equation}

Here, the angular components are transformed using trigonometric functions to ensure accurate representation of distances in angular space. To enable direct comparison with \citet{Sugar2017MPS}, the GEO vector is not normalized, in contrast to the LUTAB and ORBIT vectors.

%To ensure that all features contribute equally to the clustering process, we normalize the components of the vectors to have a mean of 0 and a variance of 1. This process standardizes each feature by centering it around zero and scaling it based on its variability, ensuring that differences in the numerical ranges of features do not bias the clustering results. Each feature is processed independently, with its mean subtracted and then divided by its standard deviation. 

\subsection{Principal Component Analysis} \label{sec:pca}

We apply PCA to evaluate whether any vector component dominates the dataset and to assess the consistency of the data \citep{pearson1901pca}. By transforming the data into a new set of orthogonal axes, PCA allows us to identify patterns and determine how different components contribute to the overall variance. The analysis focuses on the first two principal components, which capture the largest amount of variance in the dataset. Mathematically, PCA relies on the eigen decomposition of the covariance matrix of the data. The principal components correspond to the eigenvectors of this matrix, and their associated eigenvalues indicate the proportion of variance explained by each component. The first principal component is the direction along which the data varies the most, while the second principal component is orthogonal to the first and captures the next highest variance. After checking the consistency of the dataset, we apply HDBSCAN.

\subsection{HDBSCAN algorithm} \label{sec:hdbscan}

HDBSCAN is an advanced clustering algorithm that extends the capabilities of DBSCAN by introducing hierarchical clustering to handle datasets with varying density distributions. Unlike DBSCAN, which requires a fixed density threshold for defining clusters, HDBSCAN eliminates this constraint by constructing a hierarchy of clusters based on varying density levels. The algorithm begins by converting the dataset into a MST of mutual reachability distances, a measure that accounts for both point-to-point distances and local density variations. This MST is then used to create a hierarchical representation of clusters, where dense regions are identified at multiple levels of granularity. HDBSCAN employs just one parameter that defines the minimum cluster size and incorporates a stability-based metric to extract the most reliable clusters from the hierarchy. Additionally, the algorithm assigns soft membership probabilities to points, allowing it to effectively identify noise and outliers. In our study, we evaluate the performance of HDBSCAN by systematically varying the minimum sample parameter over a range from 10 to 1000. Smaller values allow the detection of finer, more localized clusters but may increase susceptibility to noise, whereas larger values yield more stable and generalized clusters, potentially overlooking smaller structures. By exploring this range, we aim to identify parameter settings that are most self-consistent according to internal clustering metrics and best reproduce the CAMS classification.

We assess the two main cluster selection methods, \textit{eom} (Excess of Mass) and \textit{leaf}. The \textit{eom} method identifies clusters by prioritizing stability across varying density levels in the hierarchical tree, selecting clusters that maximize the overall excess of mass, which corresponds to the stability of the cluster. This approach tends to produce fewer, larger clusters that are robust to noise and represent stable, well-defined groupings. In contrast, the \textit{leaf} method focuses on selecting clusters from the terminal nodes of the hierarchical tree, capturing the most localized, high-density clusters without considering the broader stability of parent clusters. This results in a greater number of smaller, granular clusters, providing finer differentiation but potentially at the expense of robustness and noise tolerance. By systematically comparing these two approaches, we aim to assess how the choice of cluster selection method affects the identification of meteor showers, particularly in datasets with complex density distributions and a high proportion of sporadic meteors. In all cases we use the standardised Euclidean
($d_{\text{sEuc}}$) distance metric, defined as:

\begin{equation} \label{eq:sEuc}
d_{\text{sEuc}}(\mathbf{x},\mathbf{y})
   = \sqrt{\sum_{i=1}^{n}\frac{(x_i - y_i)^2}{s_i^{2}}}\,,
\end{equation}

where $\mathbf{x}$ and $\mathbf{y}$ are the vectors being compared and
$s_i^{2}$ is the variance of the $i$-th feature in the sample.
Eq. \ref{eq:sEuc} is algebraically identical to the quadratic form
$(\mathbf{x}-\mathbf{y})^{\mathsf T}\mathbf{D}^{-1}(\mathbf{x}-\mathbf{y})$ with $\mathbf{D}=\mathrm{diag}(s_1^{2},\ldots,s_n^{2})$.
Because $\mathbf{D}$ is symmetric positive-definite, it admits a Cholesky factorisation $\mathbf{D}^{-1}=\mathbf{L}^{\mathsf T}\mathbf{L}$ for some invertible matrix~$\mathbf{L}$. The linear change of variables
\(
\mathbf{z} = \mathbf{L}\mathbf{x}
\)
maps each feature vector to a whitened space in which \(d_{\text{sEuc}}(\mathbf{x},\mathbf{y})
   = \left\| \mathbf{z}_x - \mathbf{z}_y \right\|_2\),
the ordinary Euclidean norm. Bijective linear maps preserve the metric axioms, so \(d_{\text{sEuc}}\) satisfies positivity, symmetry, identity of indiscernibles, and the triangle inequality \citep{DezaDeza2009}. This result holds for the LUTAB, GEO, and ORBIT vectors because every component retains non-zero variance; consequently, the clustering operates in a formally metric space.

HDBSCAN relies on an efficient nearest-neighbor search to construct a mutual reachability graph, which is fundamental for hierarchical density-based clustering. The Borůvka-Ball Tree algorithm is used to accelerate the computation of the MST and the mutual reachability distances, which are required for defining the hierarchical clustering structure. 

\subsubsection{Borůvka algorithm} \label{sec:boruvka}

Given a dataset \( X = \{x_1, x_2, \dots, x_N\} \), the core distance of a point \( x_i \), denoted as \( d_{\text{core}}(x_i) \), is defined as the distance to its \( k \)-th nearest neighbor:

\begin{equation}
d_{\text{core}}(x_i) = d_{\text{sEuc}}(x_i, x_{(k)}),
\end{equation}

where \( x_{(k)} \) is the \( k \)-th nearest neighbor of \( x_i \) based on the distance metric \( d_{\text{sEuc}}(x_i, x_j) \). The mutual reachability distance between two points \( x_i \) and \( x_j \) is then given by:

\begin{equation}
d_{\text{mr}}(x_i, x_j) = \max \left( d_{\text{core}}(x_i), d_{\text{core}}(x_j), d_{\text{sEuc}}(x_i, x_j) \right).
\end{equation}

This transformation ensures that all edges in the clustering graph respect local density constraints, preventing direct clustering of points in low-density regions.

HDBSCAN constructs a MST using Borůvka’s algorithm \cite{NESETRIL20013}. The MST is a subgraph \( T = (V, E) \) of the mutual reachability graph \( G = (V, E') \) such that: (i) \( T \) is connected and contains all vertices \( V \), (ii) \( T \) has no cycles, and (iii) the sum of edge weights is minimized. Borůvka’s algorithm iteratively finds the lightest edge (i.e., the smallest-weight edge) for each component and merges the connected components until only one remains. Mathematically, the MST is obtained as:

\begin{equation}
T = \arg \min_{T' \subseteq G} \sum_{(x_i, x_j) \in T'} d_{\text{mr}}(x_i, x_j)
\end{equation}

where \( T' \) is any spanning tree of \( G \).

\subsubsection{Ball Tree structure} \label{sec:balltree}

The computation of \( d_{\text{core}}(x_i) \) requires identifying the \( k \)-th nearest neighbor for each point. Instead of using a brute-force approach (which has \( O(N^2) \) complexity), HDBSCAN employs the Ball Tree data structure to accelerate nearest-neighbor queries.

A Ball Tree recursively partitions the dataset into a hierarchical structure of nested hyperspheres (balls), such that:
\begin{equation}
B = \{ x \in X \mid d(x, c) \leq r \}
\end{equation}
where \( c \) is the center of the ball and \( r \) is its radius. 

HDBSCAN integrates Borůvka’s algorithm with the Ball Tree data structure to enhance computational efficiency in clustering. The process begins with computing the \( k \)-nearest neighbors for each data point using the Ball Tree structure, which enables fast retrieval of neighbors and is used to determine the core distance \( d_{\text{core}}(x_i) \). Once the core distances are established, the mutual reachability graph is constructed, where edge weights between points \( x_i \) and \( x_j \) are defined by the mutual reachability distance \( d_{\text{mr}}(x_i, x_j) \). To obtain a hierarchical representation of the data, Borůvka’s algorithm is employed to efficiently compute the MST of the mutual reachability graph. The hierarchical clustering process is then performed by progressively cutting the MST, identifying clusters based on density and stability criteria. This integration of Borůvka’s MST construction with Ball Tree-based nearest-neighbor searches ensures that HDBSCAN scales effectively, maintaining both accuracy and computational efficiency for large datasets.

\subsection{Metric scores} \label{sec:metrics}

\subsubsection{Silhouette score} \label{sec:silhouette}

To evaluate the quality of the different clustering results, including that of CAMS, we compute the \textit{Silhouette score} ($S$), a standard internal validation metric that measures how well each sample fits within its assigned cluster compared to other clusters \citep{Rousseeuw1987}. For each observation \( i \), the \textit{Silhouette value} \( S(i) \) is defined as \citep{Brock2008}:

\begin{equation} \label{equation_9}
    S(i) = \frac{b(i) - a(i)}{\max(a(i), b(i))},
\end{equation}

where \( a(i) \) is the average distance between observation \( i \) and all other observations in the same cluster \( C(i) \):

\begin{equation}
    a(i) = \frac{1}{n(C(i))} \sum_{j \in C(i)} \mathrm{dist}(i, j),
\end{equation}
  
and \( b(i) \) is the smallest average distance between observation \( i \) and all observations in any other cluster \( C_k \), i.e., the nearest neighboring cluster:
  
\begin{equation}
    b(i) = \min_{C_k \in \mathcal{C} \setminus C(i)} \frac{1}{n(C_k)} \sum_{j \in C_k} \mathrm{dist}(i, j).
\end{equation}

Here, \( \mathcal{C} \) denotes the set of all clusters, \( n(C) \) is the number of elements in cluster \( C \), and \( \mathrm{dist}(i, j) \) is the pairwise distance between observations \( i \) and \( j \).

The silhouette value \( S(i) \) lies in the interval \([-1, 1]\). A value close to 1 indicates that the observation is well clustered; values near 0 suggest overlapping clusters; and negative values indicate potential misclassifications. The overall silhouette score is given by the mean of \( S(i) \) over all observations and provides a summary measure of clustering compactness and separation.

To balance computational efficiency with accuracy, we computed the Silhouette score using a random subsample of 50,000 meteors drawn from the full dataset of 316,235. This approach was necessary due to the quadratic scaling of computational cost with dataset size, as the Silhouette algorithm requires pairwise distance calculations both within and between clusters. We verified that the scores computed on the 50,000-sample subset are consistent with those obtained from the full dataset, with agreement up to the second decimal place.

To balance computational efficiency with accuracy, a sample size of 50,000 was used during the computation. This decision was motivated by the quadratic relationship between dataset size and the computational cost of the Silhouette score, as the algorithm involves pairwise distance calculations within and between clusters, making it computationally prohibitive for larger datasets. We validated that this sample size produces values consistent with those obtained using the entire dataset, agreeing up to the second decimal place.

\subsubsection{Normalized mutual information score} \label{sec:nmi}

For quantifying the agreement between the HDBSCAN clustering outputs and the meteoroid stream assignments provided by CAMS with the look-up table, we employ the Normalized Mutual Information score (NMI) \citep{Vinh2009}. NMI evaluates the similarity between two clusterings by measuring the amount of shared information while accounting for cluster size variations. The measure is based on Mutual Information (MI), which quantifies the information shared between two clusterings, \( U \) and \( V \), and is defined as:

\begin{equation}
MI(U, V) = \sum_{i} \sum_{j} P(i, j) \log \frac{P(i, j)}{P(i) P(j)},
\end{equation}

where: \( P(i, j) \) is the joint probability that a randomly chosen data point belongs to cluster \( U_i \) in clustering \( U \) and to cluster \( V_j \) in clustering \( V \), and \( P(i) \) and \( P(j) \) are the marginal probabilities of a data point belonging to cluster \( U_i \) and \( V_j \), respectively..

To normalize MI, we use the so-called entropy $H$, which quantifies the uncertainty (or information content) of a clustering:

\begin{equation}
H(U) = -\sum_{i} P(i) \log P(i)
\end{equation}
and
\begin{equation}
H(V) = -\sum_{j} P(j) \log P(j),
\end{equation}

where \( H(U) \) and \( H(V) \) represent the entropies of the two clusterings. We then normalize MI using the arithmetic mean of the entropies:

\begin{equation}
NMI(U, V) = 2 \cdot \frac{MI(U, V)}{H(U) + H(V)}.
\end{equation}

This metric measures the global alignment between the predicted clusters and the reference clusters by quantifying how much information about the reference clustering can be inferred from the predicted clustering. Derived from mutual information in information theory, NMI provides an overall assessment of structural similarity rather than focusing on fine-grained pairwise relationships. NMI values range from 0 to 1, where 1 indicates perfect alignment of the cluster structures, and 0 represents complete independence between the two clusterings. NMI is particularly robust to differences in the granularity of clusterings, making it valuable when the number of clusters varies between the predicted and reference clusters. %This normalization ensures that NMI ranges between 0 and 1, where 0 indicates no mutual information (completely independent clusterings) and 1 indicates perfect agreement. By employing this measure, we obtain a robust, label-independent evaluation of clustering performance.
 
\subsubsection{Hungarian algorithm} \label{sec:hungarian}

The final step in our clustering analysis involves aligning the unsupervised clusters produced by HDBSCAN with the meteor shower labels provided by CAMS. Since there is no inherent one-to-one correspondence between these two label sets, we employ the Hungarian algorithm (also known as the Munkres assignment algorithm) to determine an optimal mapping \citep{Kuhn1955, Kuhn1956}.

In our implementation, a contingency matrix is first constructed, where each element \( M_{ij} \) represents the number of data points that are simultaneously assigned to HDBSCAN cluster \( i \) and the CAMS label \( j \). To transform this problem into a minimization one, we define a cost matrix \( C \) as:
\[
C_{ij} = \max(M) - M_{ij},
\]
where \( \max(M) \) is the maximum value in the contingency matrix. The Hungarian algorithm then finds an assignment \( A^* \) that minimizes the total cost:
\[
A^* = \arg \min_A \sum_{(i, j) \in A} C_{ij}.
\]

Additionally, our implementation handles cases where some predicted labels remain unmapped by refining the mapping to ensure that any overlapping or unassigned labels are appropriately relabeled. This comprehensive approach ensures that the resulting mapping optimally aligns the HDBSCAN clusters with the CAMS classifications.

\subsubsection{F1 score} \label{sec:f1_score}

After obtaining the optimal mapping between HDBSCAN clusters and CAMS-labeled meteor showers using the Hungarian algorithm, we evaluate the quality of these assignments using the F1 score. The F1 score is a widely used metric in classification tasks that balances precision and recall, providing a single measure of clustering performance.

For each assigned cluster-label pair \((C_i, L_j)\), we define: (i) True Positives (TP) as the number of points correctly assigned to \( L_j \) within \( C_i \), (ii) False Positives (FP) as the number of points in \( C_i \) that do not belong to \( L_j \), and (iii) False Negatives (FN) as the number of points in \( L_j \) that were not assigned to \( C_i \). F1 is defined as:

\begin{equation}
F1 = 2 \cdot \frac{TP}{2TP + FP + FN}.
\end{equation}

The F1 score ranges from 0 to 1, where a value close to 1 indicates a strong match between the detected and reference clusters, while lower values indicate poor alignment.

%-------------------------------------------------------------------
\section{Results and discussion}

\begin{figure*}[h!]
    \centering
    \includegraphics[width=1\textwidth]{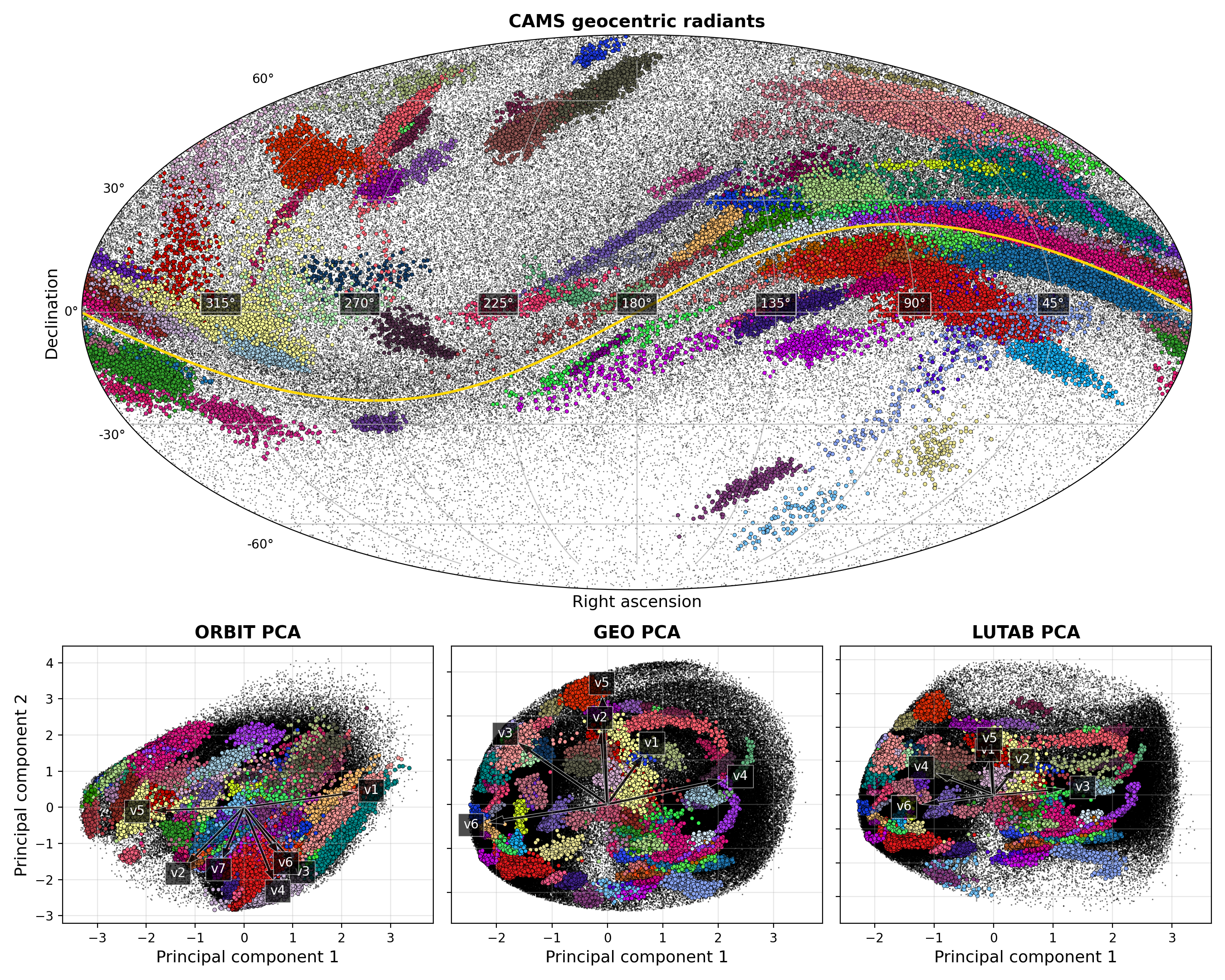}
    \caption{The top row displays the geocentric radiant distribution in a Mollweide projection, where right ascension is mapped along the horizontal axis and declination along the vertical axis, with the ecliptic plane shown as a solid yellow line. The bottom row presents the principal component analysis results for meteor showers using the ORBIT (left), GEO (center), and LUTAB (right) vectors. Arrows indicate the loadings of the original vector components (v) of ORBIT, GEO and LUTAB, illustrating their influence on the first two principal components. In both rows, meteors are colored according to their CAMS meteor shower classification, while sporadic events are represented by smaller black points. GEO is represented here with Z-score normalization as well.} 
    \label{fig:radiants_pca}
\end{figure*}

To understand the complexity of the classification problem faced in this work, we first examine the distribution of meteor radiants as seen from the Earth. Figure \ref{fig:radiants_pca} (top) presents the geocentric radiant distribution for all filtered events, color-coded according to their assigned CAMS meteor shower classification. Each meteoroid stream is represented by a unique color, consistent across all figures in this work, while sporadic meteors are depicted as small black circles. 

The figure highlights the significant overlap between meteor showers. Some showers appear diffuse, indicative of older streams that have undergone dispersion over time, while younger showers exhibit more concentrated distributions. Additionally, the dataset contains a considerable sporadic background, which accounts for approximately 70\% of the total events. This high level of noise makes the accurate identification of meteoroid streams particularly challenging. While incorporating an additional velocity dimension would improve the clustering of showers, some overlap would still persist. As expected, many streams are concentrated near the ecliptic (solid yellow line), reflecting the general inclination distribution of the Solar System's meteoroid population. The dataset also exhibits a clear norther-hemisphere bias, although this does not affect our methodology.

A complementary perspective on the classification complexity, albeit with less direct physical meaning, is provided in Figure \ref{fig:radiants_pca} (bottom). The three panels display the results of a PCA applied to the ORBIT, GEO, and LUTAB vector representations, respectively. Each plot is color-coded according to the predefined CAMS classification, providing insight into the distribution of showers within the transformed feature space. The analysis highlights the contributions of individual vector components ($v_i$ based on Eq. \ref{eqn:LUTAB}, \ref{eqn:ORBIT}, and \ref{eqn:GEO}) to the first two principal components. GEO is represented using Z-score normalization to ensure comparable feature scales, although HDBSCAN was applied to the non-normalized LUTAB vector (see Section \ref{sec:geo}).

The PCA of the ORBIT vector shows a high degree of dispersion in the distribution of meteor showers, with clusters appearing overlapped and less compact. The first component $v1$ ($q$) exhibits the highest loading, closely aligned with the first principal component. $v5$ $\cos(\omega)$) lies along the same axis but in the opposite direction, while $v2$ ($e$) and $v3$ ($i$) are nearly horizontal, also contributing to the first principal component. The remaining components—$v4$, $v6$, and $v7$—are oriented toward the negative direction of the second principal component, with $v6$ and $v7$ (the sine and cosine of $\Omega$) showing the smallest contributions, suggesting a relatively minor influence on the variance captured. In contrast, for the PCA of the GEO vector components $v2$ ($\sin(\lambda_{\odot})$) and $v5$ ($\sin(\beta_{\mathrm{g}})$) are strongly aligned with the positive second principal component axis, indicating dominant contributions to that direction. Meanwhile, $v4$ and $v6$ (cosine and scaled $V_{\mathrm{g}}$) are nearly horizontal and of similar magnitude but point in opposite directions, reflecting some redundancy. Component $v1$ ($\cos(\lambda_{\odot})$) is by far the weakest contributor. The LUTAB vector exhibits a structure broadly consistent with GEO. Components $v1$ and $v5$ ($\sin(\lambda_{\odot})$ and $\beta_{\mathrm{g}}$, respectively) show strong alignment in the direction of the second principal component, while $v3$ and $v6$ ($\sin(\lambda_{\mathrm{g}} - \lambda_{\odot})$ and $V_{\mathrm{g}}$) are aligned along the first principal component. Component $v2$ ($\cos(\lambda_{\odot})$) has the lowest loading, similar to $v1$ in GEO.

\begin{figure}%[h!]
    \centering
    \includegraphics[width=1\columnwidth]{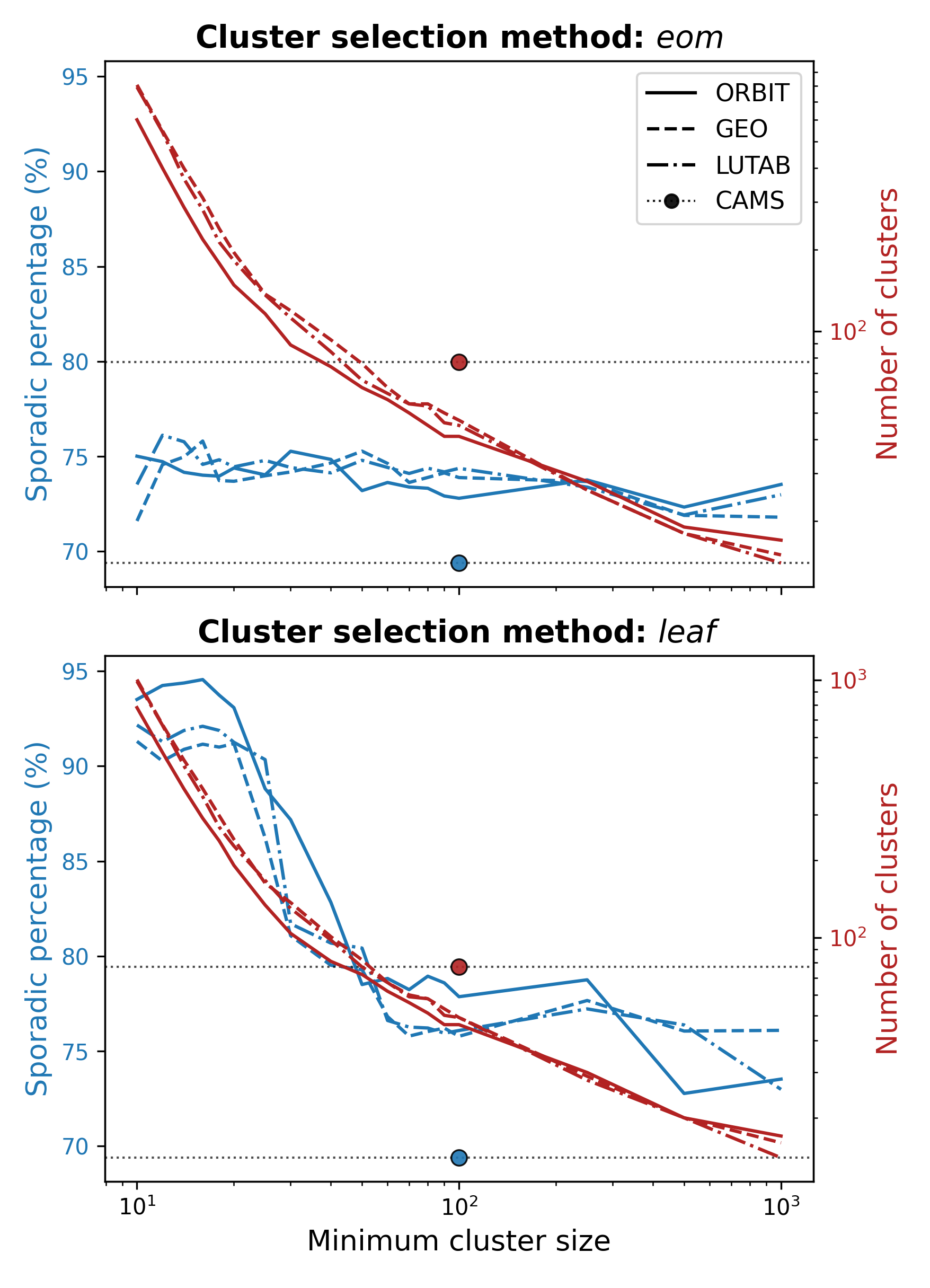}
    \caption{Sporadic percentage (blue lines) and the number of clusters (red lines) obtained from the HDBSCAN clustering algorithm as a function of the minimum cluster size for the (top) \textit{eom} and (bottom) \textit{leaf} selection methods. Solid lines represent the ORBIT vector, dashed lines correspond to GEO, and dash-dotted lines indicate LUTAB. For reference, CAMS values are shown as dots with horizontal dotted lines for minimum cluster size of 100.
}
    \label{fig:min_cluster_variation}
\end{figure}

Figure \ref{fig:min_cluster_variation} illustrates the relationship between the minimum cluster size and the number of clusters identified by HDBSCAN for both the \textit{eom} (top) and \textit{leaf} (bottom) selection methods. In both cases, the number of clusters (red lines) decreases exponentially as the minimum cluster size increases. This trend is consistent across all three vector representations. A similar behavior is observed for events classified as sporadic (blue lines) by HDBSCAN when using the \textit{leaf} selection method. However, this is not the case with \textit{eom}, where the amount of data labeled as noise appears to be almost independent of the chosen minimum cluster size.

It is possible to match the number of clusters classified by CAMS (red dots in Fig. \ref{fig:min_cluster_variation}) using any of the three vectors, but only by allowing clusters with a smaller number of members (30--50 instead of 100). The blue dots indicates the percentage of the dataset identified as sporadic meteors by CAMS—both of which are comparable. None of the HDBSCAN configurations reproduces exactly the same percentage of sporadic events as the reference classification, with the closest match obtained at the largest minimum cluster size. However, on average, HDBSCAN assigns approximately $<$5\% more events to the sporadic class. For the same minimum cluster size, HDBSCAN consistently detects fewer clusters than CAMS, regardless of the vector representation or clustering method. Conversely, the fraction of events classified as sporadic by HDBSCAN is generally higher than that reported by CAMS, except in the case of the ORBIT vector under the \textit{eom} selection method.

\begin{figure}%[h!]
    \centering
    \includegraphics[width=1\columnwidth]{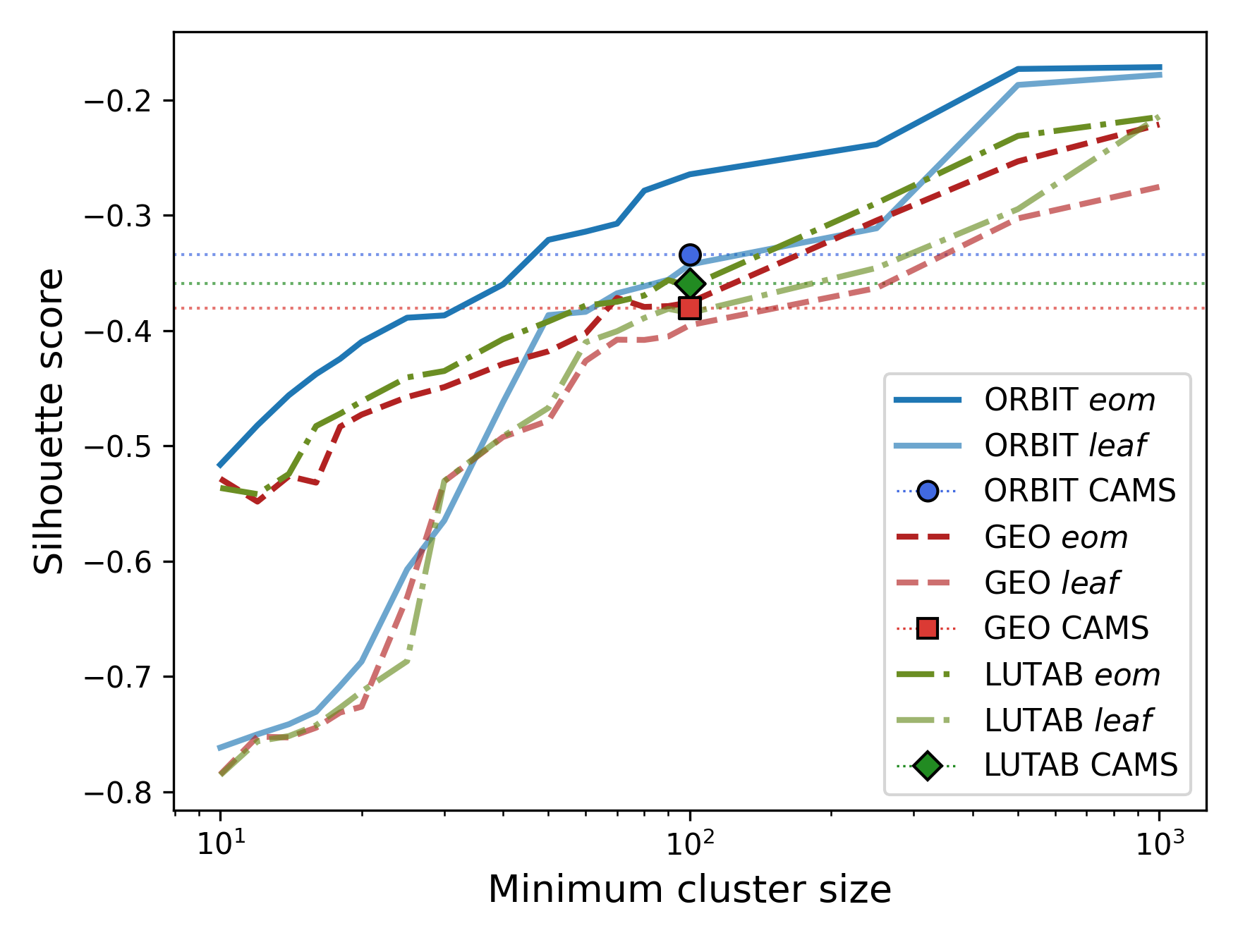}
    \caption{Silhouette score as a function of the minimum cluster size. Results are shown for the ORBIT (solid blue), GEO (dashed red), and LUTAB (dash-dotted green) vectors. Darker shades correspond to the \textit{eom} selection method, while lighter shades indicate the \textit{leaf} method. Markers denote the results at minimum cluster size of 100 for each vector: ORBIT (dot), GEO (square), and LUTAB (diamond), with accompanying horizontal lines for reference.
}
    \label{fig:silhouette_scores}
\end{figure}

Figure \ref{fig:silhouette_scores} presents the Silhouette score as a function of the minimum cluster size for the three vector representations: ORBIT (solid blue), GEO (dashed red), and LUTAB (dash-dotted green). Darker shades correspond to the \textit{eom} selection method, while lighter shades indicate \textit{leaf}. The markers denote the results for a minimum cluster size of 100, with horizontal reference lines indicating the corresponding scores.

Four key observations emerge from this figure. First, the \textit{eom} selection method consistently yields higher Silhouette scores than \textit{leaf}, regardless of the vector used. Second, among all configurations, the ORBIT vector combined with \textit{eom} achieves the highest Silhouette scores in all cases, indicating that this representation provides the most coherent cluster structures. Third, increasing the minimum cluster size generally improves the Silhouette score, suggesting that larger clusters tend to be more internally consistent and well-separated. The Silhouette score of the CAMS classification with a minimum cluster size of 100 closely matches the scores obtained using HDBSCAN with the GEO and LUTAB vectors, particularly under the \textit{eom} method, indicating consistency between the approaches. In contrast, for the ORBIT vector, HDBSCAN with \textit{eom} yields a significantly higher Silhouette score than CAMS.

Negative Silhouette scores indicate that many points are closer to a neighboring cluster than to their assigned one, suggesting poor separation between groups. This is expected given the intrinsic complexity of meteoroid stream distributions. As shown in Figure \ref{fig:radiants_pca}, many meteor showers exhibit spatial overlap and diffuse structures that complicate clear separation. Additionally, the high fraction of sporadic meteors adds ambiguity, as some inevitably coincide with or lie near existing clusters due to probabilistic reasons. These factors degrade clustering performance and lead to negative Silhouette values. Moreover, meteoroid streams do not always form well-separated structures; older or evolved streams may gradually blend into the background population. In this context, negative Silhouette scores reflect both the limitations of hard clustering algorithms—which enforce exclusive assignments—and the continuous nature of meteoroid stream space, which evolves into a quasi-random distribution over time.

\begin{figure}%[h!]
    \centering
    \includegraphics[width=1\columnwidth]{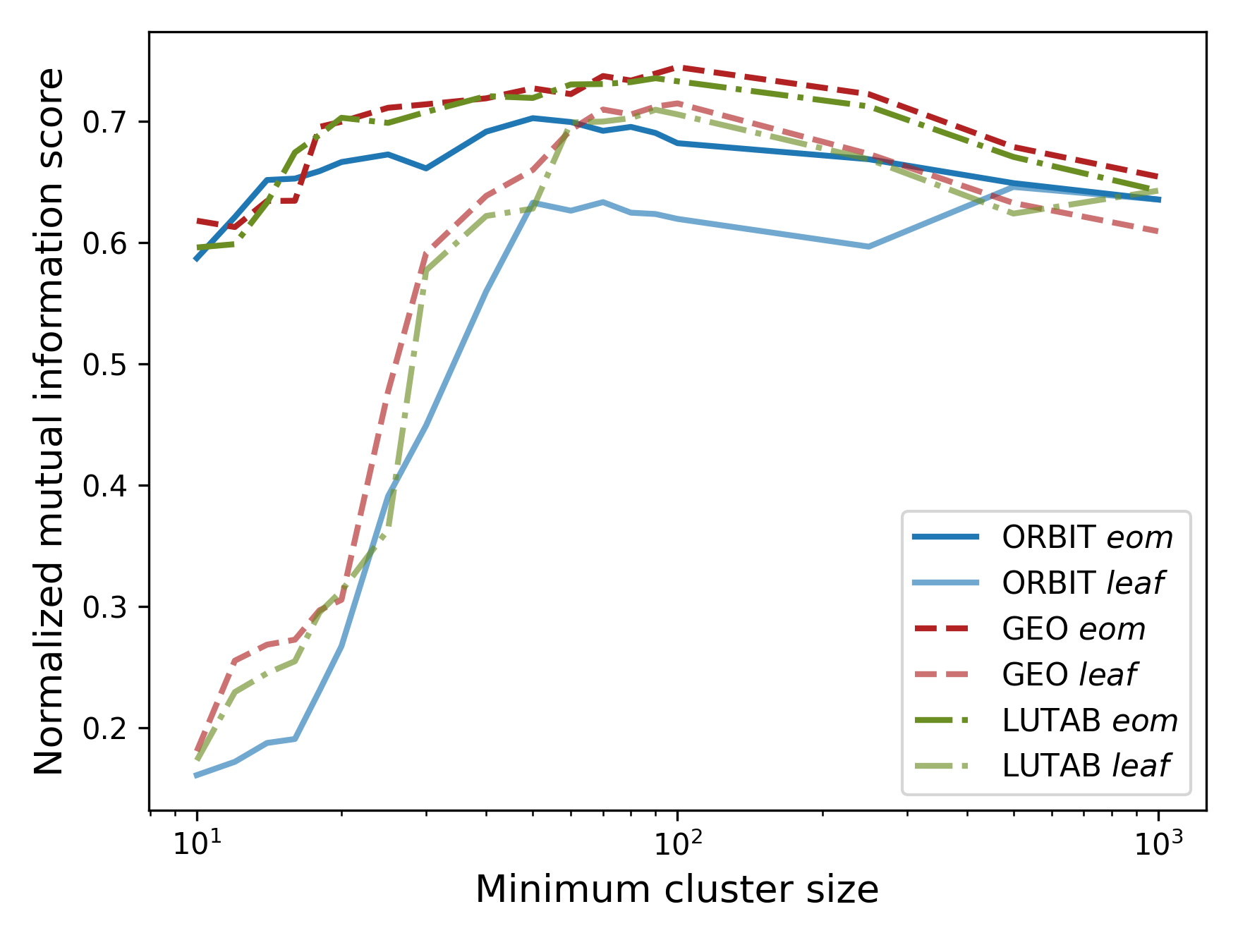}
    \caption{MNI score as a function of the minimum cluster size. HDBSCAN result is compared against CAMS classification for the ORBIT (solid blue), GEO (dashed red), and LUTAB (dash-dotted green) vectors. Darker shades correspond to the \textit{eom} selection method, while lighter shades indicate the \textit{leaf} method.}
    \label{fig:nmi_score}
\end{figure}

To further evaluate how well HDBSCAN replicates the CAMS classification, we compare the NMI score across different vector representations and clustering parameters. Figure \ref{fig:nmi_score} shows the NMI score as a function of the minimum cluster size for the ORBIT (solid blue), GEO (dashed red), and LUTAB (dash-dotted green) vectors. Darker shades represent the \textit{eom} selection method, while lighter shades correspond to \textit{leaf}. The NMI score measures the agreement between the HDBSCAN clustering results and the CAMS classification, with higher values indicating a closer match. We observe that the GEO vector with the \textit{eom} method achieves the highest similarity to the CAMS classification when HDBSCAN is allowed to identify clusters with a minimum size of 100 members, reaching an NMI score of 0.74. For a minimum cluster size of 100—the threshold used in our initial filtering—both the LUTAB and GEO vectors yield comparable results to CAMS, showing similar performances. In contrast, the ORBIT vector consistently produces the most dissimilar clustering results, with a highest NMI score of 0.68 at a minimum cluster size of 50. The \textit{leaf} method consistently performs worse in terms of similarity with the CAMS classification.

The highest overall Silhouette score, achieved using the ORBIT vector with a minimum cluster size of 1000, results in an NMI score of only 0.64, indicating a greater deviation from the CAMS classification. This suggests that while optimizing for internal cluster coherence (as measured by the Silhouette score), the clustering structure may diverge more from the predefined CAMS classification.

From this point onward, our analysis will focus on three specific cases extracted from the Silhouette and NMI scores: 1) The HDBSCAN configuration that produced the most similar result to the CAMS classification, which corresponds to the GEO vector with the \textit{eom} cluster selection method and a minimum cluster size of 100; 2) The HDBSCAN configuration that achieved the highest Silhouette score for a minimum cluster size of 100 members, which corresponds to the ORBIT vector with the \textit{eom} method; 3) The HDBSCAN configuration that yielded the global maximum Silhouette score (i.e., the highest overall Silhouette score across all tested parameter combinations), which also corresponds to the ORBIT vector with the \textit{eom} method but for a minimum cluster size of 1000 members.

\begin{figure}%[h!]
    \centering
    \includegraphics[width=1\columnwidth]{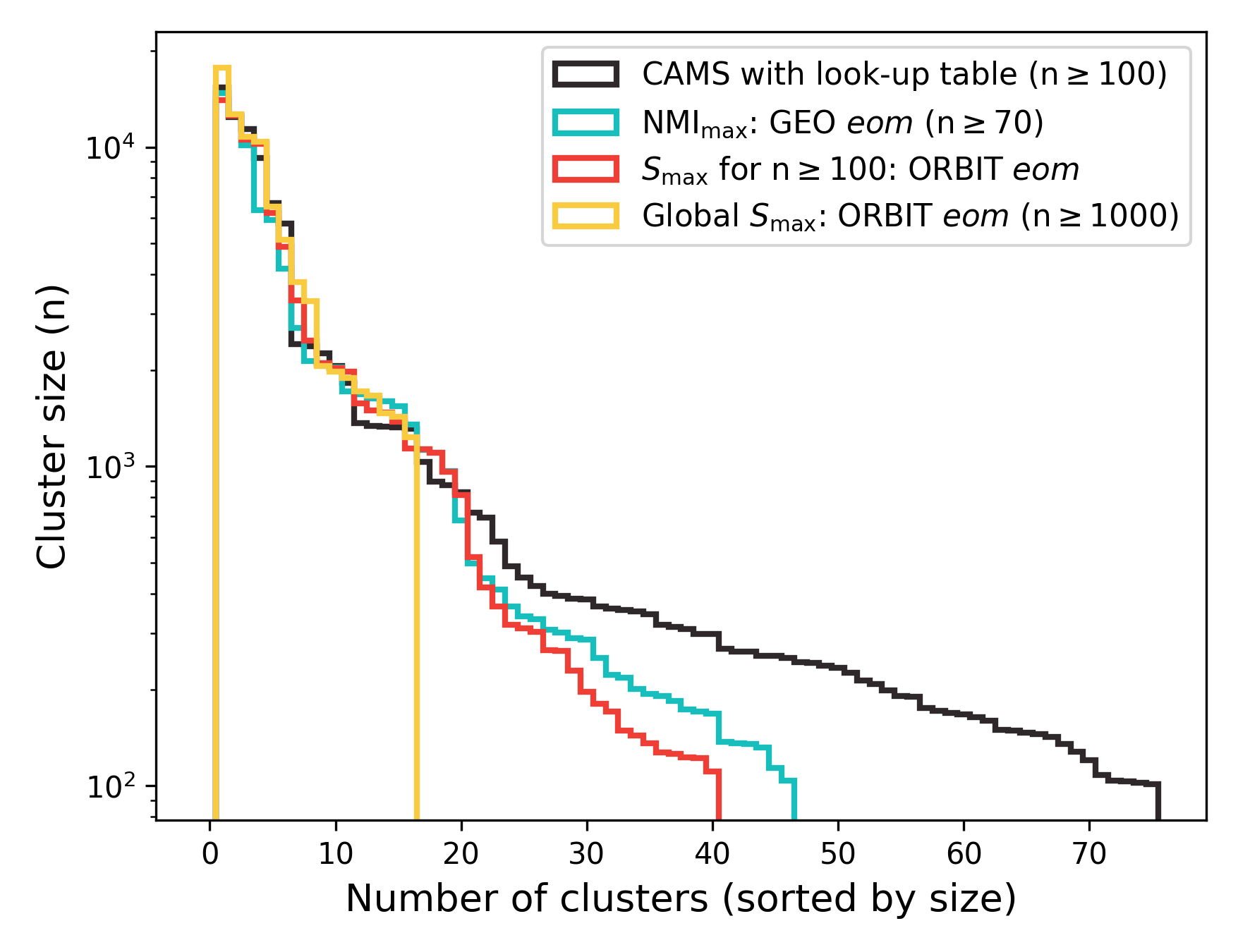}
    \caption{The x-axis represents the number of clusters, while the y-axis indicates the size of each cluster. Clusters are sorted by size in descending order, with larger clusters appearing on the left side of the x-axis. The black line represents the CAMS classification, while the cyan line corresponds to the HDBSCAN results with the maximum NMI. The red line shows the HDBSCAN results for the maximum Silhouette score at a minimum cluster size of 100, and the yellow line represents the HDBSCAN results for the overall maximum Silhouette score.
}
    \label{fig:num_clusters}
\end{figure}

Figure \ref{fig:num_clusters} provides insight into how these three cases compare in terms of cluster distribution. The x-axis represents the number of clusters, while the y-axis indicates the size of each cluster. Clusters are sorted in descending order, with the largest clusters appearing on the left. The black line represents the CAMS classification, serving as a reference. The cyan line corresponds to the HDBSCAN results that maximize the NMI score, while the red line represents the HDBSCAN results with the highest Silhouette score at a minimum cluster size of 100. Finally, the yellow line shows the results for the global maximum Silhouette score. We observe that HDBSCAN consistently identifies significantly fewer clusters than CAMS when using the look-up table, particularly for clusters with fewer than $\sim$500 members, where the divergence becomes evident. For the largest clusters, the agreement is relatively good across all vectors, with the exception of the case corresponding to the global maximum Silhouette score, which exhibits a pronounced peak at its largest clusters and a sharp drop in clusters smaller than 1236 members.

\begin{figure*}%[h!]
    \centering
    \includegraphics[width=0.8\textwidth]{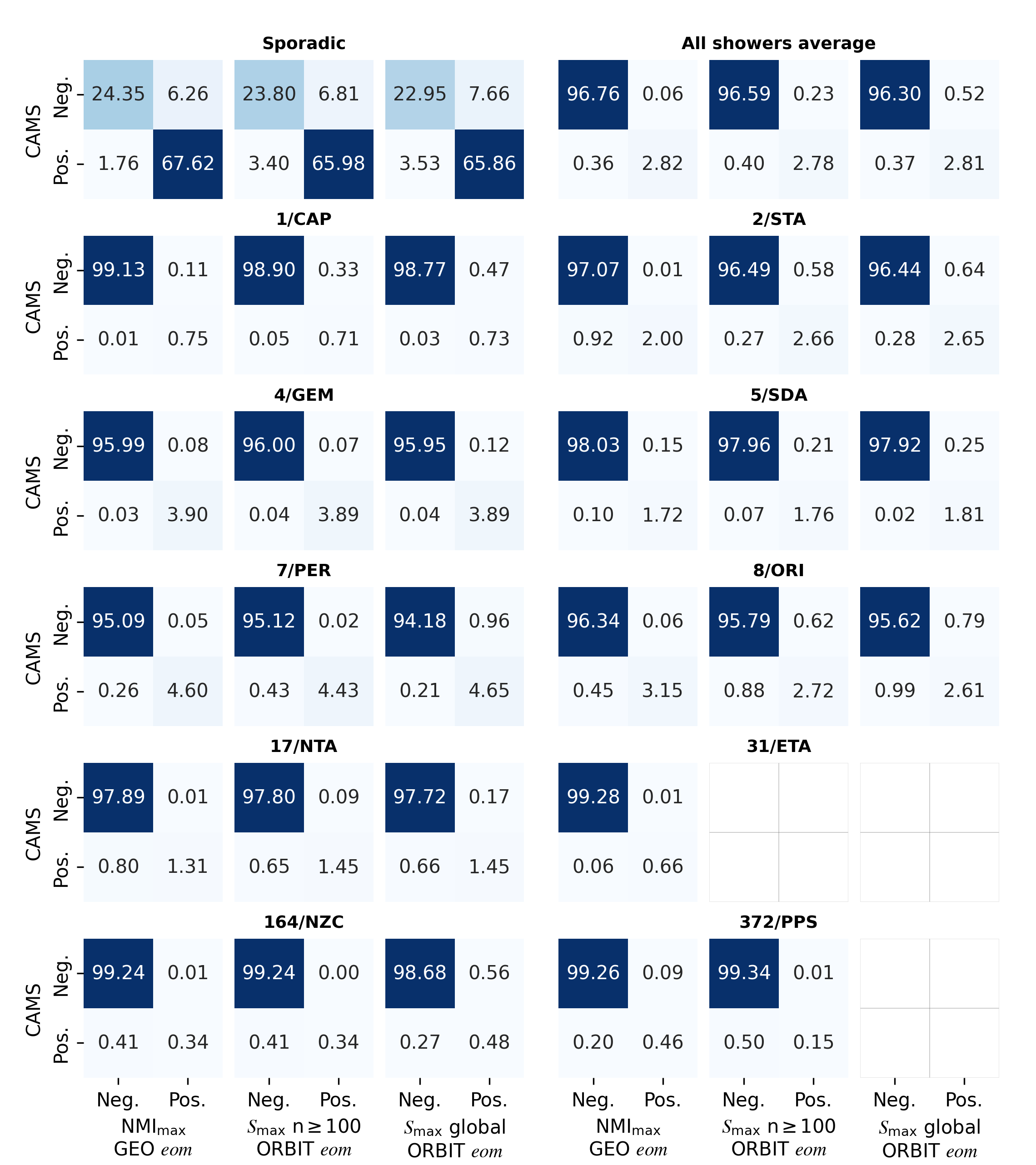}
    \caption{One-vs-rest confusion matrices constructed by comparing the meteor shower labels provided by CAMS with HDBSCAN clustering results under three conditions: maximum NMI, maximum Silhouette score at a minimum cluster size of 100, and overall maximum Silhouette score. The confusion matrix is presented for the top 10 most active meteor showers, with average performance metrics reported for these showers. Additionally, a separate confusion matrix for the sporadic background is computed using HDBSCAN's noise labels. Meteor showers are named based on official IAU identification numbers and codes. 
 }
    \label{fig:confusion_matrix}
\end{figure*}

We then determine the optimal correspondence between HDBSCAN clusters and CAMS-labeled meteor showers using the Hungarian algorithm (see Section \ref{sec:hungarian}). Based on this, Figure \ref{fig:confusion_matrix} presents the one-vs-rest confusion matrices for the top 10 most active meteor showers identified by CAMS, along with the average performance across all showers and the classification of sporadic meteors (or noise). Each shower's confusion matrix follows a one-vs-rest approach, comparing a given meteor shower against all other events, including other showers and sporadic meteors. The matrices distinguish between negative events, which CAMS did not classify as part of the shower, and positive events, which CAMS assigned to the shower. The top-left (neg.-neg.) represents meteors correctly identified as non-members, while the top-right (neg.-pos.) indicates false positives—meteors incorrectly assigned to the shower by HDBSCAN. Similarly, the bottom-left (pos.-neg.) captures false negatives, where HDBSCAN failed to classify actual shower members, and the bottom-right (pos.-pos.) corresponds to true positives, where both CAMS and HDBSCAN agree on the classification. Higher pos.-pos. values indicate better clustering performance, while high neg.-pos. or pos.-neg. values suggest classification errors.

The sporadic events show the highest level of discrepancy, reaching approximately $\sim$7--11\%. Discrepancies are consistently larger for meteors that HDBSCAN labels as noise. As expected, individual showers exhibit a very high match in the neg-neg category. For the maximum NMI case, the identification of shower members is consistently more accurate than the misclassifications, with the exception of 164/NZC. The same holds true for the maximum Silhouette score with a minimum cluster size of 100. The global maximum Silhouette score fails to identify 31/ETA and 372/PPS, whereas the maximum Silhouette score for a minimum cluster size of 100 only fails to identify 31/ETA. Overall, the average confusion matrix for all showers indicates that the classification agreement is high for the most active meteor showers.

\begin{figure*}%[h!]
    \centering
    \includegraphics[width=0.77\textwidth]{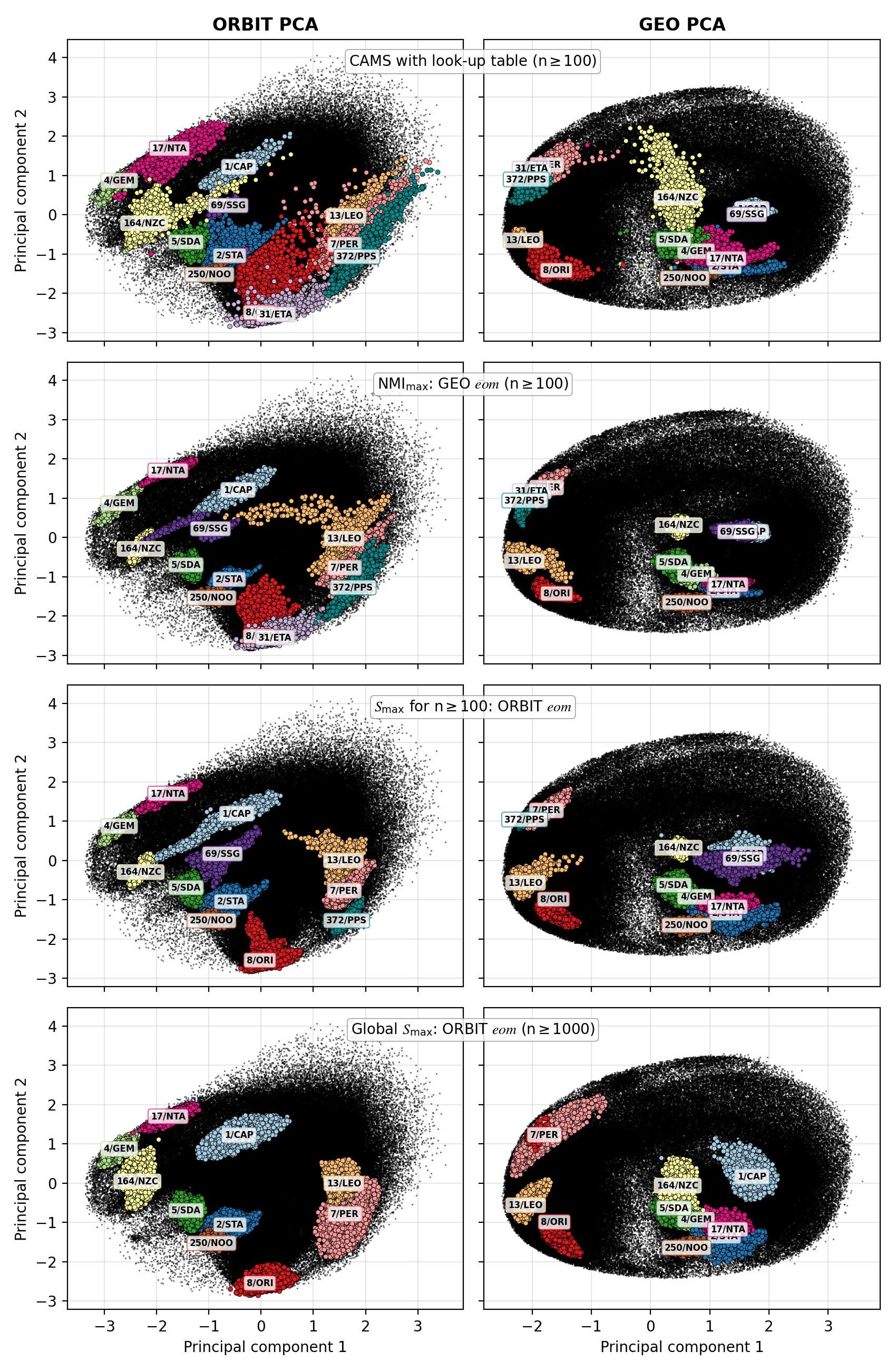}
    \caption{The columns display the principal component analysis results for the ORBIT (left) and GEO (right) vectors. The top 10 largest clusters from CAMS and each HDBSCAN result are combined (13 in total) and displayed using labels from different clustering processes across four rows for different clustering processes: CAMS original labels with a minimum cluster size of 100 (top), maximum NMI (second), maximum Silhouette score at a minimum cluster size of 100 (third), and overall maximum Silhouette score (bottom). Sporadic events in CAMS and those labeled as noise by HDBSCAN are shown as smaller black points. Meteor showers are named based on official IAU identification numbers and codes. The HDBSCAN results depicted were obtained using the \textit{eom} cluster selection method. GEO is represented here with Z-score normalization as well.}

    \label{fig:pca_clusters}
\end{figure*}

To further analyze the clustering structures, Figure \ref{fig:pca_clusters} presents the PCA results for the ORBIT (left) and GEO (right) vector representations. The figure compares the combined top 10 meteor showers from CAMS along with the clusters identified by HDBSCAN, resulting in a total of 13 displayed clusters. The figure is organized into four rows corresponding to different clustering approaches: the original CAMS classification with a minimum cluster size of 100 (top row), the HDBSCAN configuration that maximizes NMI (second row), the maximum Silhouette score with a minimum cluster size of 100 (third row), and the overall maximum Silhouette score (bottom row). Sporadic events from CAMS and those labeled as noise by HDBSCAN are represented as smaller black points, illustrating the distribution of unclustered meteors. Each cluster is labeled according to the classification method used, allowing for a direct comparison of how different clustering techniques organize the meteoroid streams. The PCA projections provide insight into the separability and compactness of clusters across different feature spaces, highlighting structural similarities and discrepancies between CAMS and HDBSCAN classifications.

In general, when comparing the PCA results for ORBIT and GEO, regardless of the classification method, the GEO vector consistently exhibits more compact clusters with less dispersion and overlap when projected onto the first two principal components. In GEO, the two meteoroid streams composing the Taurid complex (2/STA and 17/NTA) are clearly related, whereas in ORBIT, they appear far from each other together. In the second row of Figure \ref{fig:pca_clusters}, which corresponds to the clustering configuration that maximizes NMI, the clusters in GEO are not only more concentrated but also smaller in size compared to ORBIT. In contrast, for the case of maximum Silhouette score with at least 100 and 1000 members (third and fourth rows), some clusters appear more compact in the ORBIT vector. 

\begin{figure*}%[h!]
    \centering
    \includegraphics[width=0.75\textwidth]{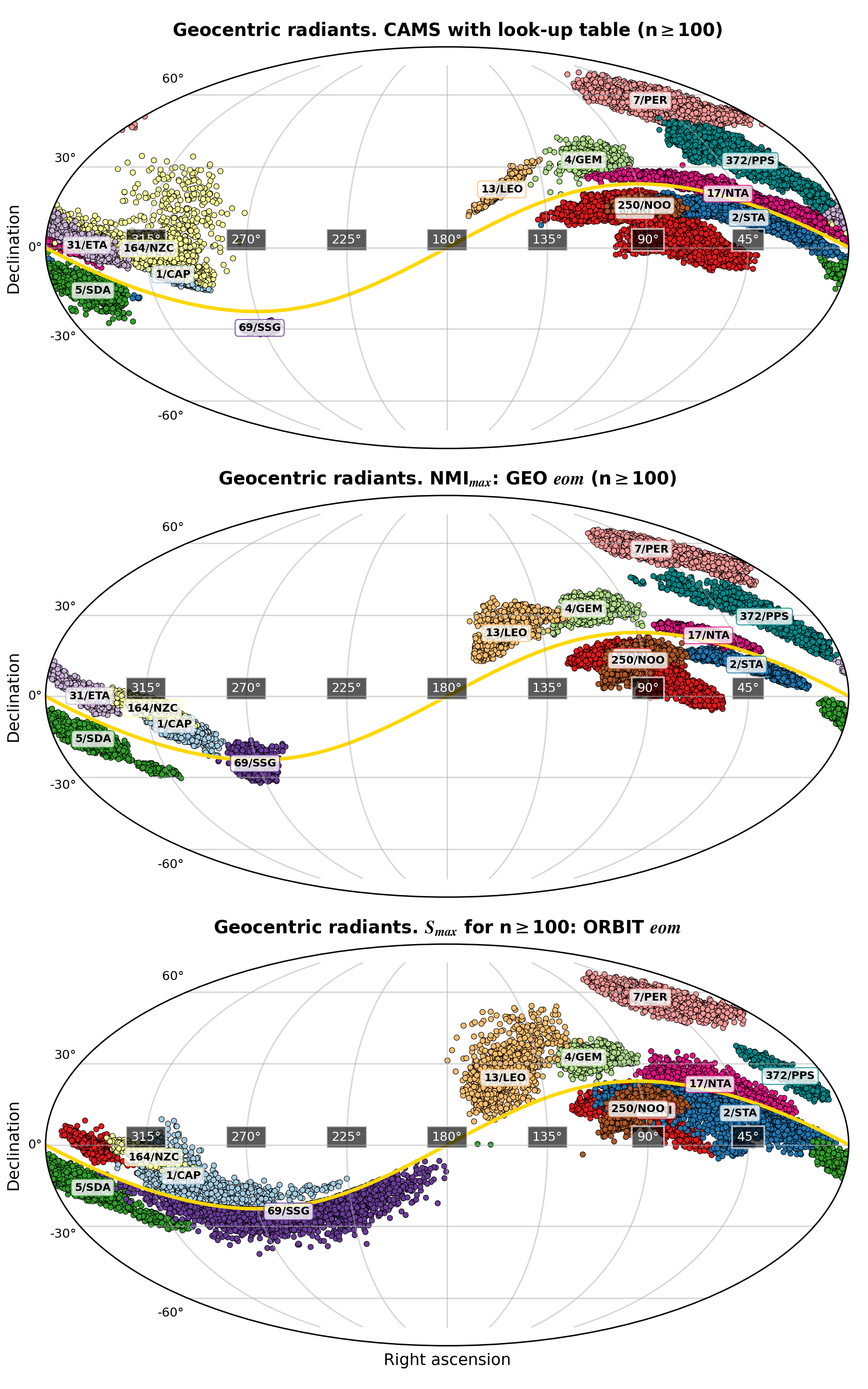}
    \caption{Geocentric radiant distributions in a Mollweide projection, where right ascension is mapped along the horizontal axis and declination along the vertical axis, and the ecliptic plane represented by a solid yellow line. The top 10 largest clusters from CAMS and each HDBSCAN result are combined (13 in total) and displayed using labels from different clustering processes across three rows: CAMS original labels with a minimum cluster size of 100 (top), maximum NMI (center), and maximum Silhouette score at a minimum cluster size of 100 (bottom). Meteor showers are named based on official IAU identification numbers and codes. The HDBSCAN results depicted were obtained using the \textit{eom} cluster selection method.
}
    \label{fig:radiant_topshowers}
\end{figure*}

For better visualization, Figure \ref{fig:radiant_topshowers} presents the geocentric radiants of the top 10 clusters from CAMS, along with the additional top 10 clusters identified by HDBSCAN, resulting in a total of 13 clusters. The figure is organized into three rows corresponding to different classification methods: (top) CAMS labels for showers with more than 100 members, (middle) HDBSCAN using the GEO vector, which maximizes NMI for clusters with at least 100 members, and (bottom) HDBSCAN using the ORBIT vector, which achieves the maximum Silhouette score for clusters with a minimum of 100 members. One striking feature is the dispersion of 8/ORI and 164/NZC as labeled by CAMS (top panel), which is not present in the HDBSCAN classification. The rest of the showers show good agreement when compared to the maximum NMI case (middle panel), with the only notable difference being the elongation of 5/SDA, 17/NTAm and 69/SSG. These elongations are even more pronounced  the maximum Silhouette score case for a minimum cluster size of 100 (bottom panel), while 8/ORI exhibits two well separate radiant subclusters. We will assess this discrepancy later in this paper. One observation is that 372/PPS is the largest only in the case of maximum NMI.

To assess the classification performance of HDBSCAN compared to CAMS, Table \ref{tab:f1_scores} presents the F1 scores for different clusters under two clustering conditions: maximum NMI and maximum Silhouette score at a minimum cluster size of 100. The table is divided into three sections, listing meteor shower codes alongside their corresponding F1 scores for each clustering approach. A dash indicates cases where no corresponding cluster was identified for a given meteor shower. The first row (Spo.) corresponds to sporadic meteors or events labeled as noise by HDBSCAN. The table highlights key differences in the number of clusters retrieved by each approach. The maximum NMI configuration (GEO) identifies a total of 47 clusters (including noise), missing 30 clusters compared to CAMS. In contrast, the maximum Silhouette score configuration with ORBIT and a minimum cluster size of 100 detects 41 clusters, missing 36 clusters from CAMS. The global maximum Silhouette score configuration (ORBIT, with at least 1000 members) retrieves only 17 clusters, failing to classify 60 clusters from CAMS. Performance varies significantly between configurations. The GEO maximum NMI case confirms 39 clusters with an F1 score above 0.5, and among them, 21 exceed 0.8, indicating strong agreement with CAMS. The ORBIT maximum Silhouette score with at least 100 members, however, confirms 30 clusters with F1 $>$ 0.5, and just 13 surpass 0.8, suggesting that this clustering approach results in fewer well-matched showers. This comparison underscores the trade-offs between clustering compactness and agreement with predefined classifications, illustrating that while optimizing for cluster structure (Silhouette score) may yield internally coherent clusters, it does not necessarily align with the CAMS classification using the look-up table. Clustering by HDBSCAN is statistically more consistent and internally coherent, but they do not necessarily have a physical meaning. For comparison, \citet{Sugar2017MPS} applied DBSCAN to a dataset of 25,885 meteors and identified 25 strong clusters and 6 weak ones, all of which correspond well to known meteoroid showers.

%Sugar: "By clustering based on solar longitude, geocentric velocity, and Sun-centered ecliptic radiant, DBSCAN efficiently identifies meteor showers while treating sporadic meteors as noise. This method successfully detected 25 known showers, proving its reliability and adaptability for large-scale datasets."

\begin{table*}%[h]
\caption{F1 scores for different clusters, evaluated against CAMS original labels with a minimum cluster size of 100 for two HDBSCAN results: maximum NMI (GEO) and maximum Silhouette score at a minimum cluster size of 100 (ORBIT). The table is divided into three groups, each listing meteor shower codes (Code) alongside their corresponding F1 scores for the two clustering processes. A dash  indicates that no cluster was identified for that meteor shower under the given condition. The first row (Spo.) corresponds to sporadic meteors or events labeled as noise by HDBSCAN. Meteor showers are named based on official IAU identification numbers and codes. The HDBSCAN results presented were obtained using the \textit{eom} cluster selection method.
}
\centering
\begin{tabular}{lcc @{\hspace{1cm}} lcc @{\hspace{1cm}} lcc}
\hline
Code & NMI$_{\mathrm{max}}$ & $S_{\mathrm{max}}$ n$\geq$100 & Code & NMI$_{\mathrm{max}}$ & $S_{\mathrm{max}}$ n$\geq$100 & Code & NMI$_{\mathrm{max}}$ & $S_{\mathrm{max}}$ n$\geq$100 \\
\hline
Spo. & 0.94 & 0.93 & 97/SCC & - & - & 390/THA & - & - \\
1/CAP & 0.92 & 0.79 & 101/PIH & - & - & 411/CAN & 0.85 & 0.78 \\
2/STA & 0.81 & 0.86 & 164/NZC & 0.62 & 0.62 & 428/DSV & 0.57 & 0.65 \\
4/GEM & 0.99 & 0.99 & 165/SZC & - & - & 456/MPS & 0.61 & 0.53 \\
5/SDA & 0.93 & 0.93 & 171/ARI & 0.92 & 0.93 & 480/TCA & - & - \\
6/LYR & 0.76 & 0.79 & 175/JPE & 0.81 & 0.74 & 494/DEL & - & - \\
7/PER & 0.97 & 0.95 & 176/PHE & 0.86 & - & 502/DRV & - & - \\
8/ORI & 0.93 & 0.78 & 183/PAU & - & - & 505/AIC & - & - \\
10/QUA & 0.94 & 0.83 & 184/GDR & - & 0.83 & 507/UAN & - & - \\
11/EVI & 0.63 & 0.51 & 187/PCA & - & - & 515/OLE & - & - \\
12/KCG & 0.56 & 0.49 & 191/ERI & 0.71 & 0.49 & 529/EHY & - & - \\
13/LEO & 0.74 & 0.70 & 197/AUD & 0.32 & - & 557/SFD & - & - \\
15/URS & 0.88 & 0.84 & 208/SPE & 0.77 & 0.34 & 581/NHE & - & - \\
16/HYD & 0.87 & 0.90 & 250/NOO & 0.75 & 0.75 & 599/POS & - & - \\
17/NTA & 0.76 & 0.80 & 253/CMI & - & - & 644/JLL & - & - \\
18/AND & 0.68 & 0.42 & 256/ORN & - & - & 648/TAL & 0.30 & - \\
19/MON & - & - & 257/ORS & 0.31 & - & 694/OMG & - & 0.62 \\
20/COM & 0.86 & 0.80 & 326/EPG & - & - & 704/OAN & - & - \\
21/AVB & 0.67 & 0.52 & 331/AHY & 0.83 & 0.48 & 714/RPI & - & - \\
22/LMI & 0.83 & 0.83 & 335/XVI & - & - & 715/ACL & - & - \\
23/EGE & 0.74 & - & 336/DKD & 0.75 & 0.20 & 746/EVE & 0.84 & 0.63 \\
26/NDA & 0.86 & 0.86 & 338/OER & 0.43 & - & 749/NMV & - & - \\
31/ETA & 0.95 & - & 372/PPS & 0.76 & 0.37 & 750/SMV & - & - \\
40/ZCY & 0.85 & 0.67 & 384/OLP & - & - & 757/CCY & 0.59 & 0.39 \\
69/SSG & 0.45 & 0.14 & 386/OBC & - & - & 842/CRN & - & - \\
96/NCC & - & - & 388/CTA & 0.59 & 0.63 & & & \\
\hline
\end{tabular}
\label{tab:f1_scores}
\end{table*}

As stated above, we find several cases where HDBSCAN clustering results deviate from the predefined CAMS classification, particularly in the grouping of meteoroid streams that share orbital similarities but are classified separately in the CAMS database. An example is the showers associated with 1P/Halley, which produces two distinct meteor showers on Earth: 8/ORI and 31/ETA \citep{Egal2020AA640A58E}. The 31/ETA occurs at the descending node of the comet's orbit, while the 8/ORI is linked to its ascending node. However, HDBSCAN using the ORBIT vector merges both showers into a single cluster. In contrast, HDBSCAN with the LUTAB or GEO vector identifies 8/ORI and 31/ETA as separate showers.  

To provide a direct numerical comparison between the meteor showers identified by HDBSCAN and those predefined in CAMS using the look-up table, Table \ref{tab:streams_comp} presents the mean features of the 13 major clusters extracted from each classification. The table includes results for three clustering approaches: CAMS original labels with a minimum cluster size of 100, the HDBSCAN configuration that maximizes NMI, and the HDBSCAN configuration that maximizes the Silhouette score with a minimum cluster size of 100. The listed parameters include key orbital and observational characteristics such as solar longitude, right ascension and declination of the radiant, geocentric velocity, perihelion distance, eccentricity, inclination, argument of perihelion, and longitude of the ascending node. The minimum and maximum values of solar longitude are computed using the 1st and 99th percentiles, while the peak value is estimated using kernel density estimation with Gaussian kernels. Additionally, to further illustrate these differences, we include a visual representation of these main parameters in the form of histograms, which can be found in the Appendix. Overall, we observe excellent agreement for the most active major showers, with the most pronounced discrepancies occurring for 69/SSG, 164/NZC and 372/PPS.

\begin{table*}%[b]
    \caption{Arithmetic mean} features of the combined top 10 clusters from CAMS and each HDBSCAN result (13 in total) for three different clustering processes: CAMS original labels with a minimum cluster size of 100 (GEO; first line for each shower), maximum NMI (ORBIT; second line for each shower), and maximum Silhouette score at a minimum cluster size of 100 (ORBIT; third line for each shower). The parameters provided include solar longitude, right ascension and declination of the radiant, geocentric velocity, perihelion distance, eccentricity, inclination, argument of perihelion, and longitude of the ascending node. The minimum and maximum values of solar longitude are computed using the 1st and 99th percentiles, respectively, while the peak value is estimated using a kernel density estimation with Gaussian kernels \citep{Scott2015lh}. Meteor showers are named based on official IAU identification numbers and codes. The HDBSCAN results presented were obtained using the \textit{eom} cluster selection method.
  
    \centering
    \begin{tabular}{lcccccccccccc}
            \hline
            Code & Clustering & $\lambda_{\odot}^{\text{min}}$ & $\lambda_{\odot}^{\text{peak}}$ & $\lambda_{\odot}^{\text{max}}$ & R.A. & Decl. & $V_{\mathrm{g}}$ & $q$ & $e$ & $i$  & $\omega$ & $\Omega$ \\  
              &   & ($^\circ$) & ($^\circ$) & ($^\circ$) & ($^\circ$) & ($^\circ$) & (km/s) & (au) & & ($^\circ$) & ($^\circ$) & ($^\circ$) \\  
            \hline
1/CAP & CAMS & 107.3 & 127.1 & 141.8 & 304.1 & -9.8 & 22.9 & 0.579 & 0.773 & 7.4 & 269.2 & 124.5 \\
 & NMI$_{\mathrm{max}}$ & 102.5 & 127.0 & 140.4 & 303.6 & -10.2 & 23.0 & 0.572 & 0.775 & 7.3 & 270.0 & 123.6 \\
 & $S_{\mathrm{max}}$ n$\geq$100 & 52.2 & 125.4 & 141.9 & 299.7 & -11.2 & 23.8 & 0.545 & 0.783 & 7.0 & 272.9 & 118.7 \\
2/STA & CAMS & 162.0 & 220.7 & 260.6 & 43.3 & 10.9 & 27.3 & 0.360 & 0.811 & 5.4 & 115.6 & 31.4 \\
 & NMI$_{\mathrm{max}}$ & 185.9 & 221.8 & 243.4 & 44.1 & 11.6 & 27.7 & 0.350 & 0.818 & 5.4 & 116.2 & 32.1 \\
 & $S_{\mathrm{max}}$ n$\geq$100 & 176.6 & 221.1 & 270.2 & 47.1 & 11.6 & 27.5 & 0.362 & 0.816 & 5.7 & 114.7 & 35.5 \\
4/GEM & CAMS & 246.6 & 261.8 & 264.0 & 112.0 & 32.5 & 34.0 & 0.145 & 0.890 & 23.2 & 324.3 & 260.7 \\
 & NMI$_{\mathrm{max}}$ & 244.8 & 261.9 & 264.2 & 111.9 & 32.4 & 34.0 & 0.144 & 0.890 & 23.1 & 324.4 & 260.6 \\
 & $S_{\mathrm{max}}$ n$\geq$100 & 245.5 & 261.9 & 264.0 & 112.0 & 32.5 & 34.0 & 0.144 & 0.890 & 23.1 & 324.4 & 260.6 \\
5/SDA & CAMS & 119.2 & 127.1 & 153.4 & 342.6 & -15.6 & 39.8 & 0.086 & 0.962 & 26.8 & 150.3 & 309.6 \\
 & NMI$_{\mathrm{max}}$ & 103.8 & 127.2 & 158.4 & 342.5 & -15.6 & 39.9 & 0.086 & 0.963 & 26.8 & 150.3 & 309.7 \\
 & $S_{\mathrm{max}}$ n$\geq$100 & 100.0 & 127.2 & 161.4 & 342.6 & -15.6 & 39.8 & 0.087 & 0.962 & 26.7 & 150.2 & 309.8 \\
7/PER & CAMS & 115.1 & 140.1 & 154.3 & 46.1 & 57.3 & 58.2 & 0.940 & 0.888 & 112.6 & 148.3 & 138.1 \\
 & NMI$_{\mathrm{max}}$ & 113.2 & 140.1 & 151.2 & 45.6 & 57.2 & 58.4 & 0.943 & 0.897 & 112.7 & 148.8 & 137.8 \\
 & $S_{\mathrm{max}}$ n$\geq$100 & 117.7 & 140.1 & 150.1 & 45.9 & 57.4 & 58.5 & 0.945 & 0.905 & 112.7 & 149.2 & 138.1 \\
8/ORI & CAMS & 151.1 & 208.4 & 237.0 & 94.4 & 13.4 & 65.5 & 0.612 & 0.902 & 160.8 & 78.7 & 25.5 \\
 & NMI$_{\mathrm{max}}$ & 161.8 & 208.2 & 231.7 & 95.4 & 14.4 & 65.7 & 0.596 & 0.913 & 162.2 & 80.6 & 27.3 \\
 & $S_{\mathrm{max}}$ n$\geq$100 & 39.7 & 209.7 & 228.5 & 85.0 & 12.6 & 65.7 & 0.573 & 0.926 & 163.3 & 85.8 & 32.3 \\
13/LEO & CAMS & 217.5 & 236.0 & 253.3 & 154.4 & 21.4 & 69.9 & 0.981 & 0.842 & 162.4 & 172.7 & 236.3 \\
 & NMI$_{\mathrm{max}}$ & 194.6 & 236.3 & 251.8 & 152.0 & 23.2 & 69.2 & 0.948 & 0.836 & 160.4 & 171.2 & 233.4 \\
 & $S_{\mathrm{max}}$ n$\geq$100 & 206.4 & 236.2 & 253.8 & 152.5 & 24.7 & 69.3 & 0.975 & 0.862 & 157.9 & 175.5 & 234.3 \\
17/NTA & CAMS & 151.9 & 226.4 & 269.8 & 49.3 & 19.8 & 27.9 & 0.359 & 0.825 & 3.2 & 294.6 & 219.9 \\
 & NMI$_{\mathrm{max}}$ & 207.3 & 226.0 & 252.3 & 57.2 & 22.4 & 28.0 & 0.365 & 0.830 & 2.8 & 293.2 & 228.4 \\
 & $S_{\mathrm{max}}$ n$\geq$100 & 197.9 & 226.1 & 256.3 & 56.4 & 22.3 & 28.1 & 0.362 & 0.830 & 3.1 & 293.5 & 227.5 \\
31/ETA & CAMS & 35.7 & 46.7 & 87.9 & 341.4 & 0.9 & 65.3 & 0.595 & 0.924 & 163.2 & 98.4 & 50.4 \\
 & NMI$_{\mathrm{max}}$ & 36.7 & 46.2 & 74.3 & 340.3 & 0.2 & 65.4 & 0.587 & 0.931 & 163.5 & 97.6 & 48.8 \\
69/SSG & CAMS & 73.1 & 84.2 & 95.9 & 271.7 & -29.6 & 25.5 & 0.447 & 0.779 & 6.1 & 106.2 & 263.9 \\
 & NMI$_{\mathrm{max}}$ & 67.7 & 83.8 & 93.4 & 271.4 & -24.7 & 28.3 & 0.377 & 0.832 & 4.3 & 75.6 & 280.2 \\
 & $S_{\mathrm{max}}$ n$\geq$100 & 15.0 & 80.1 & 129.0 & 255.4 & -24.5 & 28.2 & 0.400 & 0.834 & 5.1 & 109.8 & 247.3 \\
164/NZC & CAMS & 59.6 & 99.5 & 149.1 & 313.9 & -0.3 & 36.4 & 0.143 & 0.891 & 40.2 & 326.7 & 103.4 \\
 & NMI$_{\mathrm{max}}$ & 87.5 & 102.5 & 124.9 & 312.3 & -4.4 & 38.4 & 0.117 & 0.936 & 37.6 & 326.3 & 104.2 \\
 & $S_{\mathrm{max}}$ n$\geq$100 & 85.5 & 102.8 & 124.7 & 311.9 & -4.5 & 38.4 & 0.116 & 0.936 & 37.9 & 326.5 & 103.6 \\
250/NOO & CAMS & 226.0 & 246.0 & 262.7 & 89.6 & 15.3 & 42.0 & 0.120 & 0.984 & 24.1 & 140.8 & 65.7 \\
 & NMI$_{\mathrm{max}}$ & 227.9 & 246.8 & 268.8 & 93.1 & 13.1 & 41.8 & 0.140 & 0.983 & 27.6 & 137.5 & 69.7 \\
 & $S_{\mathrm{max}}$ n$\geq$100 & 227.4 & 246.8 & 268.7 & 93.0 & 13.1 & 41.8 & 0.139 & 0.983 & 27.6 & 137.7 & 69.4 \\
372/PPS & CAMS & 88.3 & 121.1 & 148.1 & 29.4 & 32.2 & 64.9 & 0.917 & 0.800 & 145.2 & 142.2 & 118.0 \\
 & NMI$_{\mathrm{max}}$ & 86.6 & 109.4 & 149.3 & 25.0 & 29.5 & 65.6 & 0.904 & 0.840 & 147.2 & 140.1 & 112.2 \\
 & $S_{\mathrm{max}}$ n$\geq$100 & 87.5 & 102.9 & 120.1 & 16.3 & 25.3 & 66.1 & 0.887 & 0.873 & 149.1 & 136.7 & 102.2 \\
            \hline
            \end{tabular}

    \label{tab:streams_comp}
\end{table*}

%-------------------------------------------------------------------
\section{Conclusions}

The accurate identification of meteoroid streams is essential for understanding their origins, evolution, and contributions to the sporadic meteoroid background. However, classification remains a challenging task due to significant overlap between clusters and the presence of substantial noise from sporadic meteors. This challenge is even more critical for space missions such as LUMIO, where physical parameters and impact geometry of lunar impactors must be inferred from well-characterized meteor shower observations on Earth. In this context, machine learning tools offer a promising alternative to traditional methods, which often rely on subjective criteria. In this study, we applied the HDBSCAN unsupervised clustering algorithm to the CAMS Meteoroid Orbit Database v3.0. We summarize our key findings as follows:  
\begin{itemize}
    \item Meteoroid stream classification is inherently complex, with considerable overlap between clusters and a high level of background noise, making clear separations difficult.  
    \item HDBSCAN proves effective in identifying major meteoroid streams, offering a mathematically consistent classification method. Less active showers are more difficult for HDBSCAN to identify.
    \item HDBSCAN identified 46 meteoroid streams using geocentric parameters and 40 using orbital elements.
    \item Using geocentric parameters with HDBSCAN, 39 meteoroid streams can be confirmed, with 21 showing a strong match with the CAMS classification. When using orbital elements, 30 streams are identified, but only 13 exhibit a high matching score with CAMS.  
    \item Despite reducing the number of hyperparameters compared to its predecessor DBSCAN, HDBSCAN still requires careful selection of the minimum cluster size, as this parameter significantly impacts classification performance and agreement with CAMS meteor showers.  
    \item New features of meteoroid streams identified by HDBSCAN are provided and compared with the CAMS classification.  
    \item The highest agreement with CAMS is obtained when using geocentric parameters. However, the most internally consistent clustering is always achieved using orbital elements and HDBSCAN. 
    \item The excess of mass cluster selection method consistently provides the best performance and agreement with CAMS.  
    \item While HDBSCAN produces statistically more coherent clusters than the look-up table method based on the Silhouette score, their physical validity remains to be verified.  
\end{itemize}

These results demonstrate that HDBSCAN is a robust tool for meteoroid stream identification, especially for major meteor showers, but highlight the trade-offs between clustering compactness, classification accuracy, and physical interpretation. Future work should further assess the physical consistency of HDBSCAN-defined clusters to refine meteoroid stream classification methodologies. \\

\section*{Acknowledgements}
      This work was supported by the Italian Space Agency (ASI) within the LUMIO project (ASI-PoliMi agreement n. 2024-6-HH.0).
%\end{acknowledgments}

\section*{Author contribution}
    EP-A: Conceptualization, Methodology, Software, Formal analysis, Investigation, Data curation, Writing – original draft, Visualization. 
    FF: Writing – review \& editing, Funding acquisition.
    %% But authors are expected to provide more specific details, e.g. 
%%
%%SC was responsible for writing and submitting the manuscript.
%%WWM came up with the initial research concept and edited the manuscript.
%%OTS obtained the funding and edited the manuscript.
%%EBF provided the formal analysis and validation. He also edited the manuscript.
%%GEH Supervised the undergraduates, wrote the software and administers the project github and Zenodo repositories.
%%
%% Authors can use the Contributor Role Taxonomy (CRediT) at
%% for ideas on how write a good statement tailored to their needs.

%%%%%%%%%%%%%%%%%%%%%%%%%%%%%%
\bibliography{sample631}{}

\begin{thebibliography}{}
\expandafter\ifx\csname natexlab\endcsname\relax\def\natexlab#1{#1}\fi
\providecommand{\url}[1]{\href{#1}{#1}}
\providecommand{\dodoi}[1]{doi:~\href{http://doi.org/#1}{\nolinkurl{#1}}}
\providecommand{\doeprint}[1]{\href{http://ascl.net/#1}{\nolinkurl{http://ascl.net/#1}}}
\providecommand{\doarXiv}[1]{\href{https://arxiv.org/abs/#1}{\nolinkurl{https://arxiv.org/abs/#1}}}

\bibitem[{{Ashimbekova} {et~al.}(2025){Ashimbekova}, {Vaubaillon}, \& {Koten}}]{Ashimbekova2025arXiv250316157A}
{Ashimbekova}, A., {Vaubaillon}, J., \& {Koten}, P. 2025, arXiv e-prints, arXiv:2503.16157.
\newblock \doarXiv{2503.16157}

\bibitem[{{Avdellidou} \& {Vaubaillon}(2019)}]{Avdellidou2019MNRAS4845212A}
{Avdellidou}, C., \& {Vaubaillon}, J. 2019, \mnras, 484, 5212, \dodoi{10.1093/mnras/stz355}

\bibitem[{{Babadzhanov} \& {Obrubov}(1987)}]{Babadzhanov1987PAICz67141B}
{Babadzhanov}, P.~B., \& {Obrubov}, I.~V. 1987, Publications of the Astronomical Institute of the Czechoslovak Academy of Sciences, 2, 141

\bibitem[{Balducci(2024)}]{Balducci2024}
Balducci, S. 2024, PhD thesis.
\newblock \url{https://amslaurea.unibo.it/id/eprint/32544/}

\bibitem[{{Bellot Rubio} {et~al.}(1998){Bellot Rubio}, {Ortiz}, \& {Sada}}]{BellotRubio1998EMP82575B}
{Bellot Rubio}, L.~R., {Ortiz}, J.~L., \& {Sada}, P.~V. 1998, Earth Moon and Planets, 82-83, 575, \dodoi{10.1023/A:1017097724416}

\bibitem[{{Bonanos} {et~al.}(2018){Bonanos}, {Avdellidou}, {Liakos}, {Xilouris}, {Dapergolas}, {Koschny}, {Bellas-Velidis}, {Boumis}, {Charmandaris}, {Fytsilis}, \& {Maroussis}}]{Bonanos2018AA612A76B}
{Bonanos}, A.~Z., {Avdellidou}, C., {Liakos}, A., {et~al.} 2018, \aap, 612, A76, \dodoi{10.1051/0004-6361/201732109}

\bibitem[{{Bouley} {et~al.}(2012){Bouley}, {Baratoux}, {Vaubaillon}, {Mocquet}, {Le Feuvre}, {Colas}, {Benkhaldoun}, {Daassou}, {Sabil}, \& {Lognonn{\'e}}}]{Bouley2012Icar218115B}
{Bouley}, S., {Baratoux}, D., {Vaubaillon}, J., {et~al.} 2012, \icarus, 218, 115, \dodoi{10.1016/j.icarus.2011.11.028}

\bibitem[{Brock {et~al.}(2008)Brock, Pihur, Datta, \& Datta}]{Brock2008}
Brock, G., Pihur, V., Datta, S., \& Datta, S. 2008, Journal of Statistical Software, 25, 1–22, \dodoi{10.18637/jss.v025.i04}

\bibitem[{Campello {et~al.}(2013)Campello, Moulavi, \& Sander}]{Campello2013}
Campello, R. J. G.~B., Moulavi, D., \& Sander, J. 2013, in Advances in Knowledge Discovery and Data Mining, ed. J.~Pei, V.~S. Tseng, L.~Cao, H.~Motoda, \& G.~Xu (Berlin, Heidelberg: Springer Berlin Heidelberg), 160--172

\bibitem[{Campello {et~al.}(2015)Campello, Moulavi, Zimek, \& Sander}]{Campello2015}
Campello, R. J. G.~B., Moulavi, D., Zimek, A., \& Sander, J. 2015, ACM Trans. Knowl. Discov. Data, 10, \dodoi{10.1145/2733381}

\bibitem[{{Ceplecha} {et~al.}(1998){Ceplecha}, {Borovi{\v{c}}ka}, {Elford}, {Revelle}, {Hawkes}, {Porub{\v{c}}an}, \& {{\v{S}}imek}}]{Ceplecha1998SSRv}
{Ceplecha}, Z., {Borovi{\v{c}}ka}, J., {Elford}, W.~G., {et~al.} 1998, Space Science Reviews, 84, 327, \dodoi{10.1023/A:1005069928850}

\bibitem[{{Cipriano} {et~al.}(2018){Cipriano}, {Dei Tos}, \& {Topputo}}]{Cipriano2018FrASS529C}
{Cipriano}, A.~M., {Dei Tos}, D.~A., \& {Topputo}, F. 2018, Frontiers in Astronomy and Space Sciences, 5, 29, \dodoi{10.3389/fspas.2018.00029}

\bibitem[{Deza \& Deza(2009)}]{DezaDeza2009}
Deza, M.~M., \& Deza, E. 2009, Encyclopedia of Distances (Berlin, Heidelberg: Springer Berlin Heidelberg), 1--583, \dodoi{10.1007/978-3-642-00234-2_1}

\bibitem[{{Drummond}(1981)}]{Drummond1981Icar45545D}
{Drummond}, J.~D. 1981, Icarus, 45, 545, \dodoi{10.1016/0019-1035(81)90020-8}

\bibitem[{{Egal} {et~al.}(2020){Egal}, {Brown}, {Rendtel}, {Campbell-Brown}, \& {Wiegert}}]{Egal2020AA640A58E}
{Egal}, A., {Brown}, P.~G., {Rendtel}, J., {Campbell-Brown}, M., \& {Wiegert}, P. 2020, \aap, 640, A58, \dodoi{10.1051/0004-6361/202038115}

\bibitem[{{Ester} {et~al.}(1996){Ester}, {Kriegel}, {Sander}, \& {Xu}}]{1996kddmconf226E}
{Ester}, M., {Kriegel}, H.-P., {Sander}, J., \& {Xu}, X. 1996, in Second International Conference on Knowledge Discovery and Data Mining (KDD'96). Proceedings of a conference held August 2-4, ed. D.~W. {Pfitzner} \& J.~K. {Salmon}, 226--331

\bibitem[{{Galligan}(2001)}]{Galligan2001MNRAS327623G}
{Galligan}, D.~P. 2001, MNRAS, 327, 623, \dodoi{10.1046/j.1365-8711.2001.04858.x}

\bibitem[{Ghamarian \& Marquis(2019)}]{Ghamarian2019}
Ghamarian, I., \& Marquis, E. 2019, Ultramicroscopy, 200, 28, \dodoi{https://doi.org/10.1016/j.ultramic.2019.01.011}

\bibitem[{{Hajdukov{\'a}} {et~al.}(2023){Hajdukov{\'a}}, {Rudawska}, {Jopek}, {Koseki}, {Kokhirova}, \& {Neslu{\v{s}}an}}]{Hajdukova2023AA671A155H}
{Hajdukov{\'a}}, M., {Rudawska}, R., {Jopek}, T.~J., {et~al.} 2023, \aap, 671, A155, \dodoi{10.1051/0004-6361/202244964}

\bibitem[{Hemmelgarn {et~al.}(2024)Hemmelgarn, Moskovitz, Pilorz, \& Jenniskens}]{Hemmelgarn_2024}
Hemmelgarn, S., Moskovitz, N., Pilorz, S., \& Jenniskens, P. 2024, The Planetary Science Journal, 5, 242, \dodoi{10.3847/PSJ/ad8346}

\bibitem[{Ibrahim \& Shafiq(2018)}]{Ibrahim2018}
Ibrahim, R., \& Shafiq, M.~O. 2018, in 2018 Thirteenth International Conference on Digital Information Management (ICDIM), 130--135, \dodoi{10.1109/ICDIM.2018.8847135}

\bibitem[{{Jenniskens}(1998)}]{Jenniskens1998EPS50555J}
{Jenniskens}, P. 1998, Earth, Planets and Space, 50, 555, \dodoi{10.1186/BF03352149}

\bibitem[{{Jenniskens}(2006)}]{Jenniskens2006mspcbookJ}
---. 2006, {Meteor Showers and their Parent Comets} (Cambridge University Press)

\bibitem[{{Jenniskens}(2008a)}]{Jenniskens2008Icar19413J}
---. 2008a, Icarus, 194, 13, \dodoi{10.1016/j.icarus.2007.09.016}

\bibitem[{Jenniskens(2008b)}]{Jenniskens2008book}
Jenniskens, P. 2008b, Mostly Dormant Comets and their Disintegration into Meteoroid Streams: A Review, ed. J.~M. Trigo-Rodr{\'i}guez, F.~J.~M. Rietmeijer, J.~Llorca, \& D.~Janches (New York, NY: Springer New York), 505--520, \dodoi{10.1007/978-0-387-78419-9\_66}

\bibitem[{Jenniskens(2023)}]{Jenniskens2023Atlas}
---. 2023, Atlas of Earth's Meteor Showers, 1st edn. (Amsterdam: Elsevier)

\bibitem[{{Jenniskens} {et~al.}(2011){Jenniskens}, {Gural}, {Dynneson}, {Grigsby}, {Newman}, {Borden}, {Koop}, \& {Holman}}]{Jenniskens2011Icar}
{Jenniskens}, P., {Gural}, P.~S., {Dynneson}, L., {et~al.} 2011, Icarus, 216, 40, \dodoi{10.1016/j.icarus.2011.08.012}

\bibitem[{{Jenniskens} {et~al.}(2018){Jenniskens}, {Baggaley}, {Crumpton}, {Aldous}, {Pokorny}, {Janches}, {Gural}, {Samuels}, {Albers}, {Howell}, {Johannink}, {Breukers}, {Odeh}, {Moskovitz}, {Collison}, \& {Ganju}}]{Jenniskens2018PSS15421J}
{Jenniskens}, P., {Baggaley}, J., {Crumpton}, I., {et~al.} 2018, Planetary and Space Science, 154, 21, \dodoi{10.1016/j.pss.2018.02.013}

\bibitem[{{Jopek} {et~al.}(2024){Jopek}, {Neslu{\v{s}}an}, {Rudawska}, \& {Hajdukov{\'a}}}]{Jopek2024AA682A159J}
{Jopek}, T.~J., {Neslu{\v{s}}an}, L., {Rudawska}, R., \& {Hajdukov{\'a}}, M. 2024, \aap, 682, A159, \dodoi{10.1051/0004-6361/202347910}

\bibitem[{{Kholshevnikov} {et~al.}(2016){Kholshevnikov}, {Kokhirova}, {Babadzhanov}, \& {Khamroev}}]{Kholshevnikov2016MNRAS}
{Kholshevnikov}, K.~V., {Kokhirova}, G.~I., {Babadzhanov}, P.~B., \& {Khamroev}, U.~H. 2016, Monthly Notices of the Royal Astronomical Society, 462, 2275, \dodoi{10.1093/mnras/stw1712}

\bibitem[{{Koschny} \& {Borovicka}(2017)}]{Koschny2017JIMO4591K}
{Koschny}, D., \& {Borovicka}, J. 2017, WGN, Journal of the International Meteor Organization, 45, 91

\bibitem[{{Koschny} {et~al.}(2019){Koschny}, {Soja}, {Engrand}, {Flynn}, {Lasue}, {Levasseur-Regourd}, {Malaspina}, {Nakamura}, {Poppe}, {Sterken}, \& {Trigo-Rodr{\'\i}guez}}]{Koschny2019SSRv21534K}
{Koschny}, D., {Soja}, R.~H., {Engrand}, C., {et~al.} 2019, \ssr, 215, 34, \dodoi{10.1007/s11214-019-0597-7}

\bibitem[{{Koseki}(2017)}]{Koseki2017eMetN}
{Koseki}, M. 2017, eMeteorNews, 2, 105

\bibitem[{{Koseki}(2022)}]{Koseki2022eMetN}
---. 2022, eMeteorNews, 7, 381

\bibitem[{Kuhn(1955)}]{Kuhn1955}
Kuhn, H.~W. 1955, Naval Research Logistics Quarterly, 2, 83–97, \dodoi{10.1002/nav.3800020109}

\bibitem[{Kuhn(1956)}]{Kuhn1956}
---. 1956, Naval Research Logistics Quarterly, 3, 253–258, \dodoi{10.1002/nav.3800030404}

\bibitem[{Lentzakis {et~al.}(2020)Lentzakis, Seshadri, Akkinepally, Vu, \& Ben-Akiva}]{Lentzakis2020}
Lentzakis, A.~F., Seshadri, R., Akkinepally, A., Vu, V.-A., \& Ben-Akiva, M. 2020, Transportation Research Part C: Emerging Technologies, 118, 102685, \dodoi{https://doi.org/10.1016/j.trc.2020.102685}

\bibitem[{{Liakos} {et~al.}(2024){Liakos}, {Bonanos}, {Xilouris}, {Koschny}, {Bellas-Velidis}, {Boumis}, {Maroussis}, \& {Moissl}}]{Liakos2024AA687A14L}
{Liakos}, A., {Bonanos}, A.~Z., {Xilouris}, E.~M., {et~al.} 2024, \aap, 687, A14, \dodoi{10.1051/0004-6361/202449542}

\bibitem[{{Madiedo} {et~al.}(2015){Madiedo}, {Ortiz}, {Morales}, \& {Cabrera-Ca{\~n}o}}]{Madiedo2015PSS111105M}
{Madiedo}, J.~M., {Ortiz}, J.~L., {Morales}, N., \& {Cabrera-Ca{\~n}o}, J. 2015, \planss, 111, 105, \dodoi{10.1016/j.pss.2015.03.018}

\bibitem[{McInnes {et~al.}(2017)McInnes, Healy, \& Astels}]{McInnes2017}
McInnes, L., Healy, J., \& Astels, S. 2017, Journal of Open Source Software, 2, 205, \dodoi{10.21105/joss.00205}

\bibitem[{{Milanov} {et~al.}(2019){Milanov}, {Milanova}, \& {Kholshevnikov}}]{Milanov2019CeMDA}
{Milanov}, D.~V., {Milanova}, Y.~V., \& {Kholshevnikov}, K.~V. 2019, Celestial Mechanics and Dynamical Astronomy, 131, 5, \dodoi{10.1007/s10569-019-9884-6}

\bibitem[{Mishra {et~al.}(2022)Mishra, Monath, Boratko, Kobren, \& McCallum}]{Mishra2022}
Mishra, S., Monath, N., Boratko, M., Kobren, A., \& McCallum, A. 2022, Proceedings of the AAAI Conference on Artificial Intelligence, 36, 7788, \dodoi{10.1609/aaai.v36i7.20747}

\bibitem[{{Moreno-Ib{\'a}{\~n}ez} {et~al.}(2015){Moreno-Ib{\'a}{\~n}ez}, {Gritsevich}, \& {Trigo-Rodr{\'\i}guez}}]{Moreno2015Icar250544M}
{Moreno-Ib{\'a}{\~n}ez}, M., {Gritsevich}, M., \& {Trigo-Rodr{\'\i}guez}, J.~M. 2015, \icarus, 250, 544, \dodoi{10.1016/j.icarus.2014.12.027}

\bibitem[{{Murray}(1982)}]{Murray1982Icar49125M}
{Murray}, C.~D. 1982, Icarus, 49, 125, \dodoi{10.1016/0019-1035(82)90062-8}

\bibitem[{{Neslu{\v{s}}an} \& {Hajdukov{\'a}}(2017)}]{Nesluvan2017AA598A40N}
{Neslu{\v{s}}an}, L., \& {Hajdukov{\'a}}, M. 2017, \aap, 598, A40, \dodoi{10.1051/0004-6361/201629659}

\bibitem[{{Nesvorn{\'y}} {et~al.}(2010){Nesvorn{\'y}}, {Jenniskens}, {Levison}, {Bottke}, {Vokrouhlick{\'y}}, \& {Gounelle}}]{Nesvorny2010ApJ713816N}
{Nesvorn{\'y}}, D., {Jenniskens}, P., {Levison}, H.~F., {et~al.} 2010, The Astrophysical Journal, 713, 816, \dodoi{10.1088/0004-637X/713/2/816}

\bibitem[{Nešetřil {et~al.}(2001)Nešetřil, Milková, \& Nešetřilová}]{NESETRIL20013}
Nešetřil, J., Milková, E., \& Nešetřilová, H. 2001, Discrete Mathematics, 233, 3, \dodoi{https://doi.org/10.1016/S0012-365X(00)00224-7}

\bibitem[{{Ortiz} {et~al.}(2015){Ortiz}, {Madiedo}, {Morales}, {Santos-Sanz}, \& {Aceituno}}]{Ortiz2015MNRAS454344O}
{Ortiz}, J.~L., {Madiedo}, J.~M., {Morales}, N., {Santos-Sanz}, P., \& {Aceituno}, F.~J. 2015, \mnras, 454, 344, \dodoi{10.1093/mnras/stv1921}

\bibitem[{{Pauls} \& {Gladman}(2005)}]{Pauls2005MandPS40241P}
{Pauls}, A., \& {Gladman}, B. 2005, Meteoritics and Planetary Science, 40, 1241, \dodoi{10.1111/j.1945-5100.2005.tb00186.x}

\bibitem[{{Pe{\~n}a-Asensio} {et~al.}(2024{\natexlab{a}}){Pe{\~n}a-Asensio}, {K{\"u}ppers}, {Trigo-Rodr{\'\i}guez}, \& {Rimola}}]{Eloy2024PSJ5206P}
{Pe{\~n}a-Asensio}, E., {K{\"u}ppers}, M., {Trigo-Rodr{\'\i}guez}, J.~M., \& {Rimola}, A. 2024{\natexlab{a}}, Planetary Science Journal, 5, 206, \dodoi{10.3847/PSJ/ad6b0f}

\bibitem[{{Pe{\~n}a-Asensio} \& {S{\'a}nchez-Lozano}(2024)}]{Eloy2024AdSpR741073P}
{Pe{\~n}a-Asensio}, E., \& {S{\'a}nchez-Lozano}, J.~M. 2024, Advances in Space Research, 74, 1073, \dodoi{10.1016/j.asr.2024.05.005}

\bibitem[{{Pe{\~n}a-Asensio} {et~al.}(2021){Pe{\~n}a-Asensio}, {Trigo-Rodr{\'\i}guez}, {Gritsevich}, \& {Rimola}}]{Eloy2021MNRAS5044829P}
{Pe{\~n}a-Asensio}, E., {Trigo-Rodr{\'\i}guez}, J.~M., {Gritsevich}, M., \& {Rimola}, A. 2021, \mnras, 504, 4829, \dodoi{10.1093/mnras/stab999}

\bibitem[{{Pe{\~n}a-Asensio} {et~al.}(2022){Pe{\~n}a-Asensio}, {Trigo-Rodr{\'\i}guez}, \& {Rimola}}]{Eloy2022AJ16476P}
{Pe{\~n}a-Asensio}, E., {Trigo-Rodr{\'\i}guez}, J.~M., \& {Rimola}, A. 2022, \aj, 164, 76, \dodoi{10.3847/1538-3881/ac75d2}

\bibitem[{{Pe{\~n}a-Asensio} {et~al.}(2023){Pe{\~n}a-Asensio}, {Trigo-Rodr{\'\i}guez}, {Rimola}, {Corretg{\'e}-Gilart}, \& {Koschny}}]{Eloy2023MNRAS5205173P}
{Pe{\~n}a-Asensio}, E., {Trigo-Rodr{\'\i}guez}, J.~M., {Rimola}, A., {Corretg{\'e}-Gilart}, M., \& {Koschny}, D. 2023, \mnras, 520, 5173, \dodoi{10.1093/mnras/stad102}

\bibitem[{{Pe{\~n}a-Asensio} {et~al.}(2024{\natexlab{b}}){Pe{\~n}a-Asensio}, {Visuri}, {Trigo-Rodr{\'\i}guez}, {Socas-Navarro}, {Gritsevich}, {Siljama}, \& {Rimola}}]{Eloy2024Icar40815844P}
{Pe{\~n}a-Asensio}, E., {Visuri}, J., {Trigo-Rodr{\'\i}guez}, J.~M., {et~al.} 2024{\natexlab{b}}, \icarus, 408, 115844, \dodoi{10.1016/j.icarus.2023.115844}

\bibitem[{Pearson(1901)}]{pearson1901pca}
Pearson, K. 1901, The London, Edinburgh, and Dublin Philosophical Magazine and Journal of Science, 2, 559, \dodoi{10.1080/14786440109462720}

\bibitem[{Rousseeuw(1987)}]{Rousseeuw1987}
Rousseeuw, P.~J. 1987, Journal of Computational and Applied Mathematics, 20, 53, \dodoi{https://doi.org/10.1016/0377-0427(87)90125-7}

\bibitem[{Scott(2015)}]{Scott2015lh}
Scott, D.~W. 2015, Multivariate density estimation, 2nd edn., Wiley Series in Probability and Statistics (Chichester, England: Wiley-Blackwell)

\bibitem[{{Shober} {et~al.}(2025){Shober}, {Courtot}, \& {Vaubaillon}}]{Shober2025AA693A23S}
{Shober}, P.~M., {Courtot}, A., \& {Vaubaillon}, J. 2025, \aap, 693, A23, \dodoi{10.1051/0004-6361/202452123}

\bibitem[{{Shober} \& {Vaubaillon}(2024)}]{Shober2024AA686A130S}
{Shober}, P.~M., \& {Vaubaillon}, J. 2024, \aap, 686, A130, \dodoi{10.1051/0004-6361/202349024}

\bibitem[{{Silber} {et~al.}(2018){Silber}, {Boslough}, {Hocking}, {Gritsevich}, \& {Whitaker}}]{Silber2018AdSpR}
{Silber}, E.~A., {Boslough}, M., {Hocking}, W.~K., {Gritsevich}, M., \& {Whitaker}, R.~W. 2018, Advances in Space Research, 62, 489, \dodoi{10.1016/j.asr.2018.05.010}

\bibitem[{{Southworth} \& {Hawkins}(1963)}]{Southworth1963SCoA7261S}
{Southworth}, R.~B., \& {Hawkins}, G.~S. 1963, Smithsonian Contributions to Astrophysics, 7, 261

\bibitem[{{Sugar} {et~al.}(2017){Sugar}, {Moorhead}, {Brown}, \& {Cooke}}]{Sugar2017MPS}
{Sugar}, G., {Moorhead}, A., {Brown}, P., \& {Cooke}, W. 2017, Meteoritics and Planetary Science, 52, 1048, \dodoi{10.1111/maps.12856}

\bibitem[{{Suggs} {et~al.}(2008){Suggs}, {Cooke}, {Suggs}, {Swift}, \& {Hollon}}]{Suggs2008EMP102293S}
{Suggs}, R.~M., {Cooke}, W.~J., {Suggs}, R.~J., {Swift}, W.~R., \& {Hollon}, N. 2008, Earth Moon and Planets, 102, 293, \dodoi{10.1007/s11038-007-9184-0}

\bibitem[{{Topputo} {et~al.}(2023){Topputo}, {Merisio}, {Franzese}, {Giordano}, {Massari}, {Pilato}, {Labate}, {Cervone}, {Speretta}, {Menicucci}, {Turan}, {Bertels}, {Vennekens}, {Walker}, \& {Koschny}}]{Topputo2023Icar38915213T}
{Topputo}, F., {Merisio}, G., {Franzese}, V., {et~al.} 2023, Icarus, 389, 115213, \dodoi{10.1016/j.icarus.2022.115213}

\bibitem[{Tran {et~al.}(2021)Tran, Cao, \& Tran}]{Tran2021}
Tran, T.-H., Cao, T.-D., \& Tran, T.-T.-H. 2021, HDBSCAN: Evaluating the Performance of Hierarchical Clustering for Big Data, ed. N.~H. Phuong \& V.~Kreinovich (Cham: Springer International Publishing), 273--283, \dodoi{10.1007/978-3-030-76620-7_24}

\bibitem[{{Turchak} \& {Gritsevich}(2014)}]{Turchak2014JTAM44d15T}
{Turchak}, L.~I., \& {Gritsevich}, M.~I. 2014, Journal of Theoretical and Applied Mechanics, 44, 15, \dodoi{10.2478/jtam-2014-0020}

\bibitem[{{Vaubaillon} {et~al.}(2019){Vaubaillon}, {Neslu{\v{s}}an}, {Sekhar}, {Rudawska}, \& {Ryabova}}]{Vaubaillon2019msmebook161V}
{Vaubaillon}, J., {Neslu{\v{s}}an}, L., {Sekhar}, A., {Rudawska}, R., \& {Ryabova}, G.~O. 2019, in Meteoroids: Sources of Meteors on Earth and Beyond, ed. G.~O. {Ryabova}, D.~J. {Asher}, \& M.~J. {Campbell-Brown}, 161

\bibitem[{{{\v{D}}uri{\v{s}}ov{\'a}} {et~al.}(2024){{\v{D}}uri{\v{s}}ov{\'a}}, {Neslu{\v{s}}an}, {Hajdukov{\'a}}, {Rudawska}, \& {Jopek}}]{Durisova2024MNRAS5353661D}
{{\v{D}}uri{\v{s}}ov{\'a}}, S., {Neslu{\v{s}}an}, L., {Hajdukov{\'a}}, M., {Rudawska}, R., \& {Jopek}, T.~J. 2024, \mnras, 535, 3661, \dodoi{10.1093/mnras/stae2547}

\bibitem[{{Vida} {et~al.}(2023){Vida}, {Brown}, {Devillepoix}, {Wiegert}, {Moser}, {Matlovi{\v{c}}}, {Herd}, {Hill}, {Sansom}, {Towner}, {T{\'o}th}, {Cooke}, \& {Hladiuk}}]{DenisVida2023NatAs7318V}
{Vida}, D., {Brown}, P.~G., {Devillepoix}, H. A.~R., {et~al.} 2023, Nature Astronomy, 7, 318, \dodoi{10.1038/s41550-022-01844-3}

\bibitem[{Vinh {et~al.}(2009)Vinh, Epps, \& Bailey}]{Vinh2009}
Vinh, N.~X., Epps, J., \& Bailey, J. 2009, in Proceedings of the 26th Annual International Conference on Machine Learning, ICML '09 (New York, NY, USA: Association for Computing Machinery), 1073–1080, \dodoi{10.1145/1553374.1553511}

\bibitem[{Wang {et~al.}(2021)Wang, Chen, Chen, \& Mou}]{Wang2021}
Wang, L., Chen, P., Chen, L., \& Mou, J. 2021, Journal of Marine Science and Engineering, 9, \dodoi{10.3390/jmse9060566}

\bibitem[{Wang {et~al.}(2025)Wang, Yang, Zhuang, Lin, Tian, Su, \& Cheng}]{Wang2025}
Wang, S.-M., Yang, W.-R., Zhuang, Q.-Y., {et~al.} 2025, Applied Sciences, 15, \dodoi{10.3390/app15052621}

\bibitem[{{Williams} \& {Jopek}(2014)}]{Williams2014me13conf179W}
{Williams}, I.~P., \& {Jopek}, T.~J. 2014, in Meteoroids 2013, ed. T.~J. {Jopek}, F.~J.~M. {Rietmeijer}, J.~{Watanabe}, \& I.~P. {Williams}, 179

\bibitem[{{Yang} \& {Ishiguro}(2015)}]{YangIshiguro2015ApJ81387Y}
{Yang}, H., \& {Ishiguro}, M. 2015, The Astrophysical Journal, 813, 87, \dodoi{10.1088/0004-637X/813/2/87}

\bibitem[{Zhang {et~al.}(2018)Zhang, Lee, \& Lee}]{Zhang2018}
Zhang, D., Lee, K., \& Lee, I. 2018, Expert Systems with Applications, 92, 1, \dodoi{https://doi.org/10.1016/j.eswa.2017.09.040}

\end{thebibliography}
\bibliographystyle{aasjournal}

%%%%%%%%%%%%%%%%%%%%%%%%%%%%%%
\appendix

%\section{Appendix information}
Mean features of the combined top 10 largest meteor showers from CAMS and each HDBSCAN result (13 in total) for three different clustering processes: CAMS original labels with a minimum cluster size of 100 (black lines), maximum NMI (cyan lines), and maximum Silhouette score at a minimum cluster size of 100 (red lines). The parameters provided include solar longitude, right ascension and declination of the radiant, geocentric velocity, perihelion distance, eccentricity, inclination, argument of perihelion, and longitude of the ascending node. Meteor showers are named based on official IAU identification numbers and codes. The HDBSCAN results depicted were obtained using the \textit{eom} cluster selection method.

\begin{figure*}[h!]
    \centering
    \includegraphics[width=0.8\textwidth]{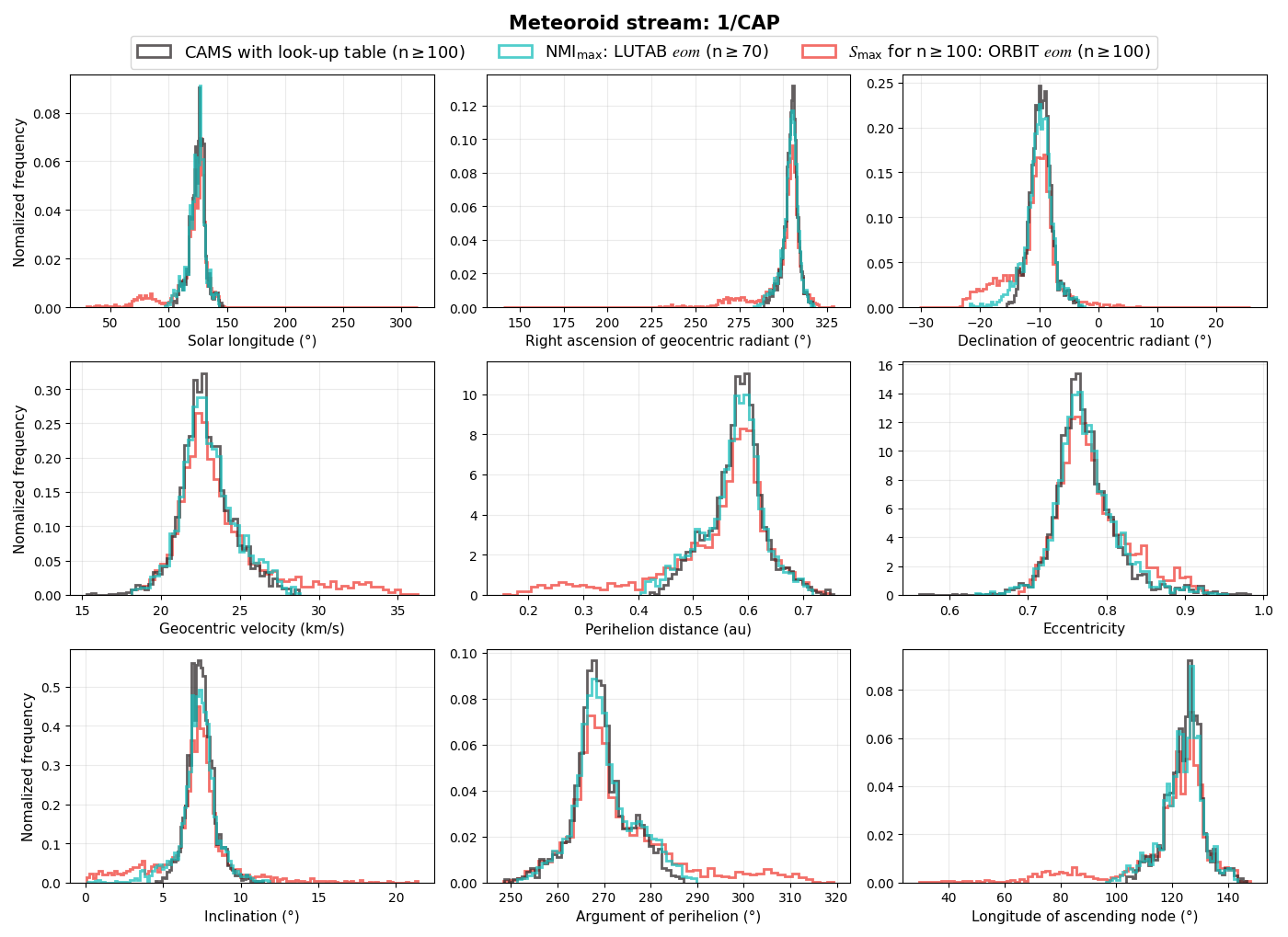}
    \caption{Mean features of the $\alpha$-Capricornids.}
    \label{fig:shower_CAP}
\end{figure*}

\begin{figure*}[h!]
    \centering
    \includegraphics[width=0.8\textwidth]{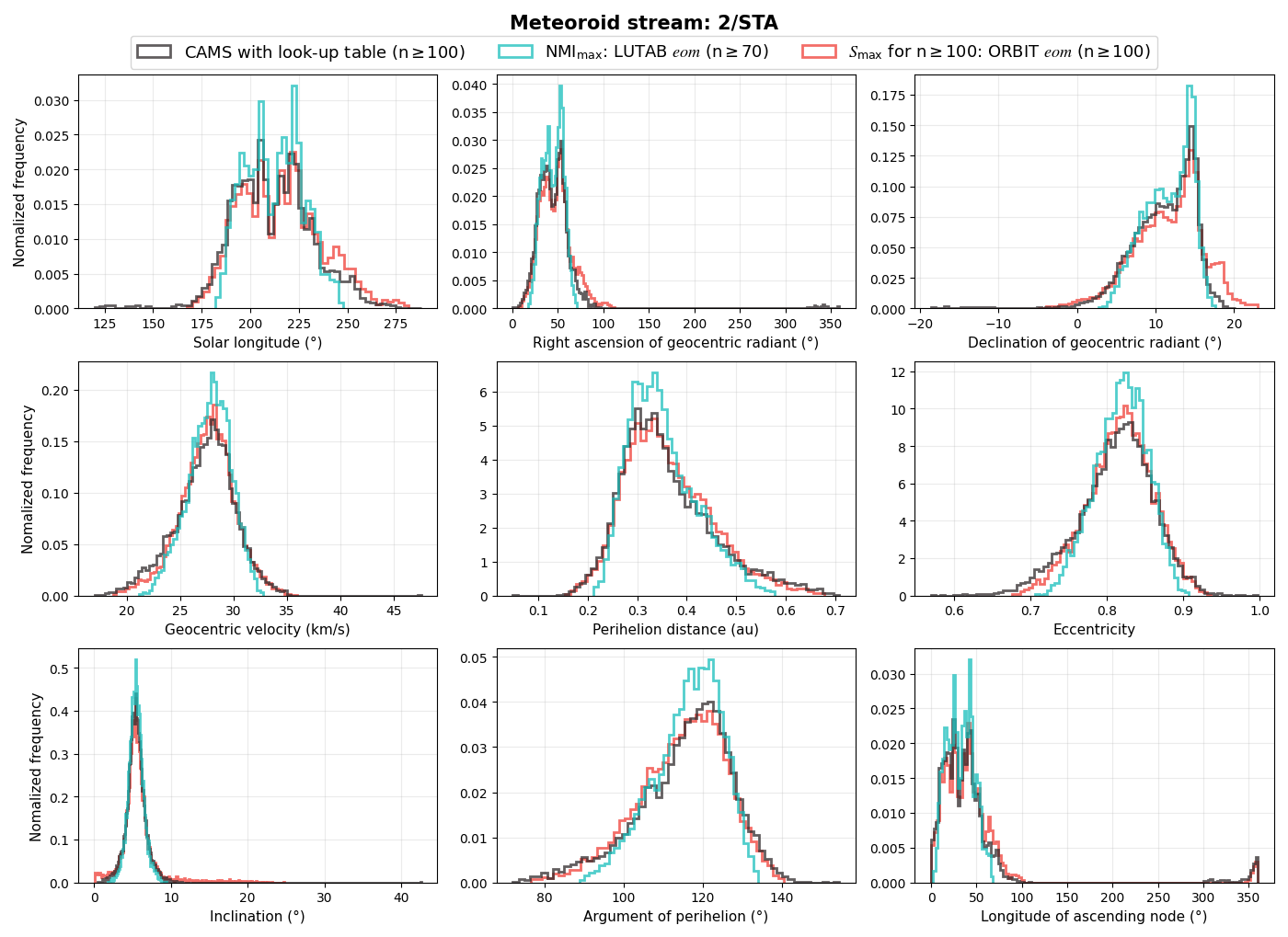}
    \caption{Mean features of the Southern Taurids.}
    \label{fig:shower_STA}
\end{figure*}

\begin{figure*}[h!]
    \centering
    \includegraphics[width=0.8\textwidth]{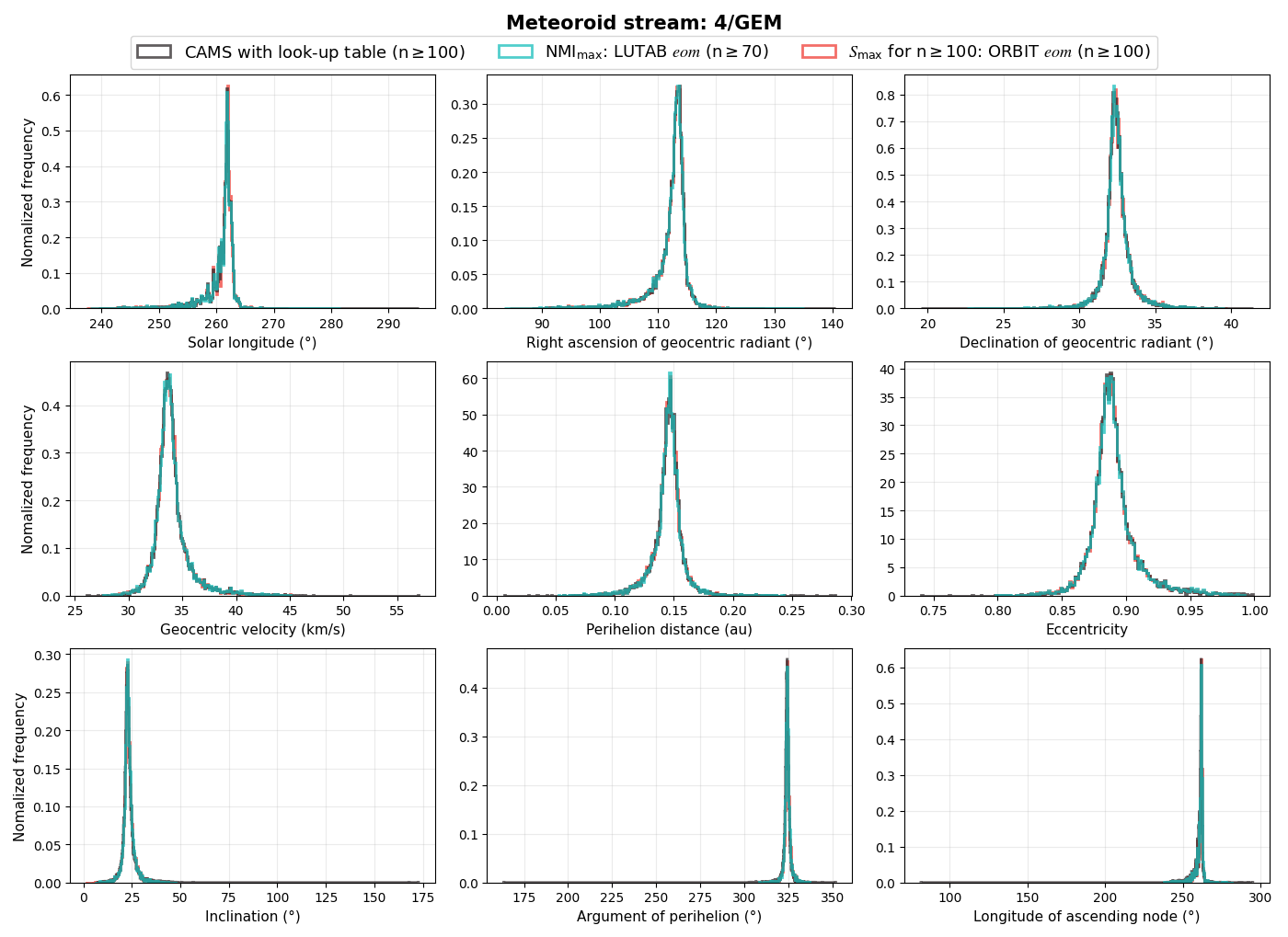}
    \caption{Mean features of the Geminids.}
    \label{fig:shower_GEM}
\end{figure*}

\begin{figure*}[h!]
    \centering
    \includegraphics[width=0.8\textwidth]{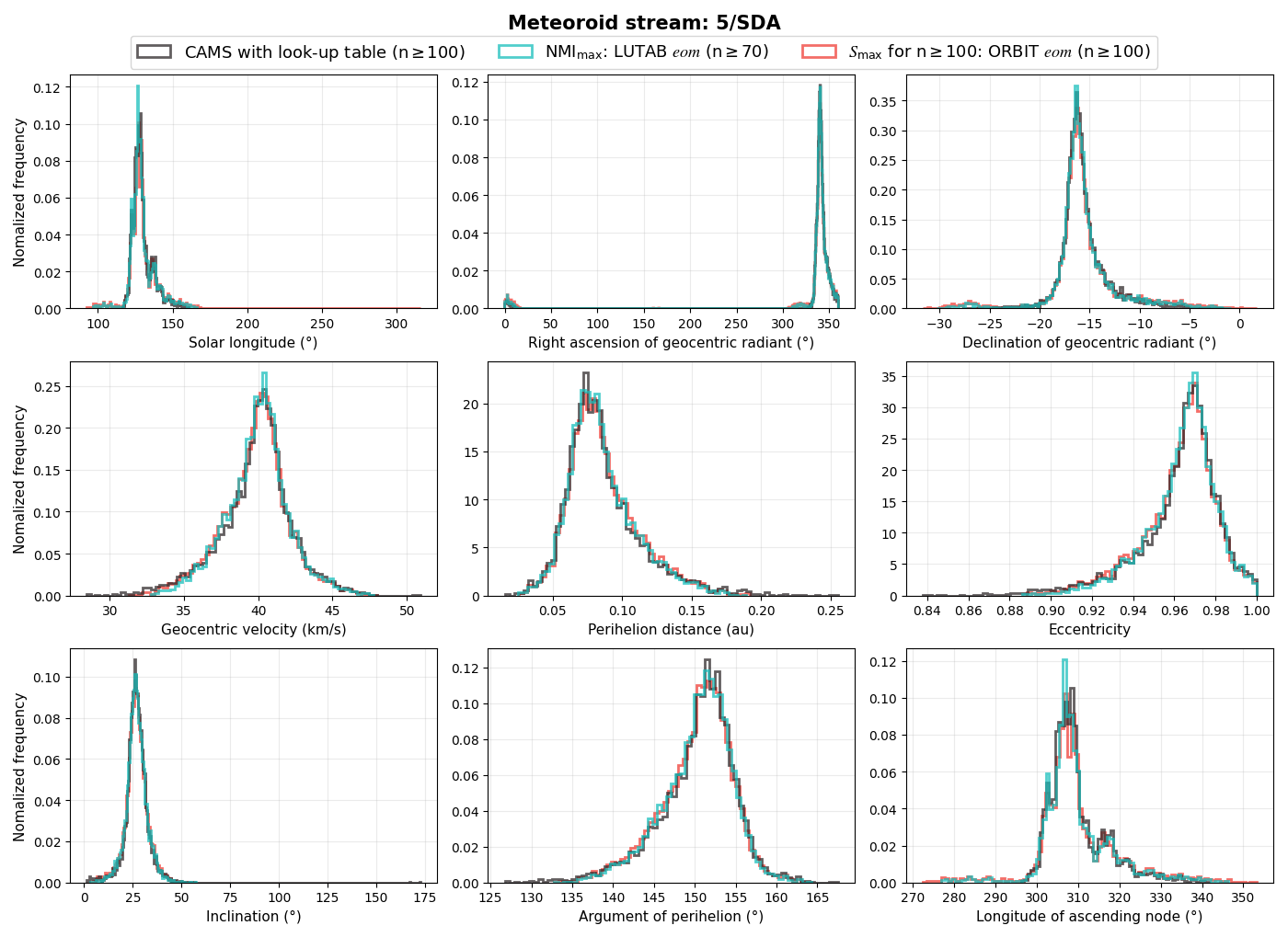}
    \caption{Mean features of the Southern $\delta$-Aquariids.}
    \label{fig:shower_SDA}
\end{figure*}

\begin{figure*}[h!]
    \centering
    \includegraphics[width=0.8\textwidth]{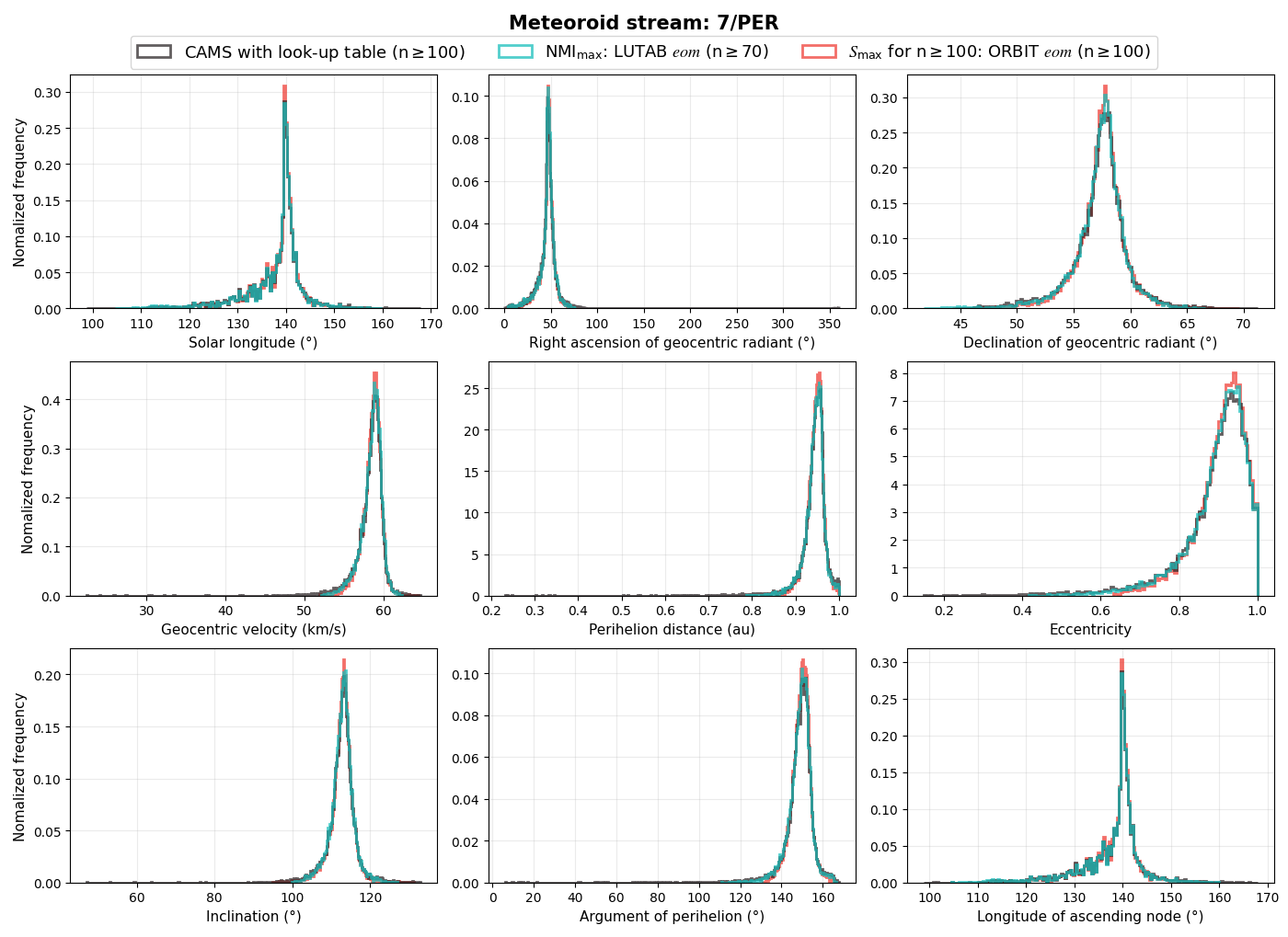}
    \caption{Mean features of the Perseids.}
    \label{fig:shower_PER}
\end{figure*}

\begin{figure*}[h!]
    \centering
    \includegraphics[width=0.8\textwidth]{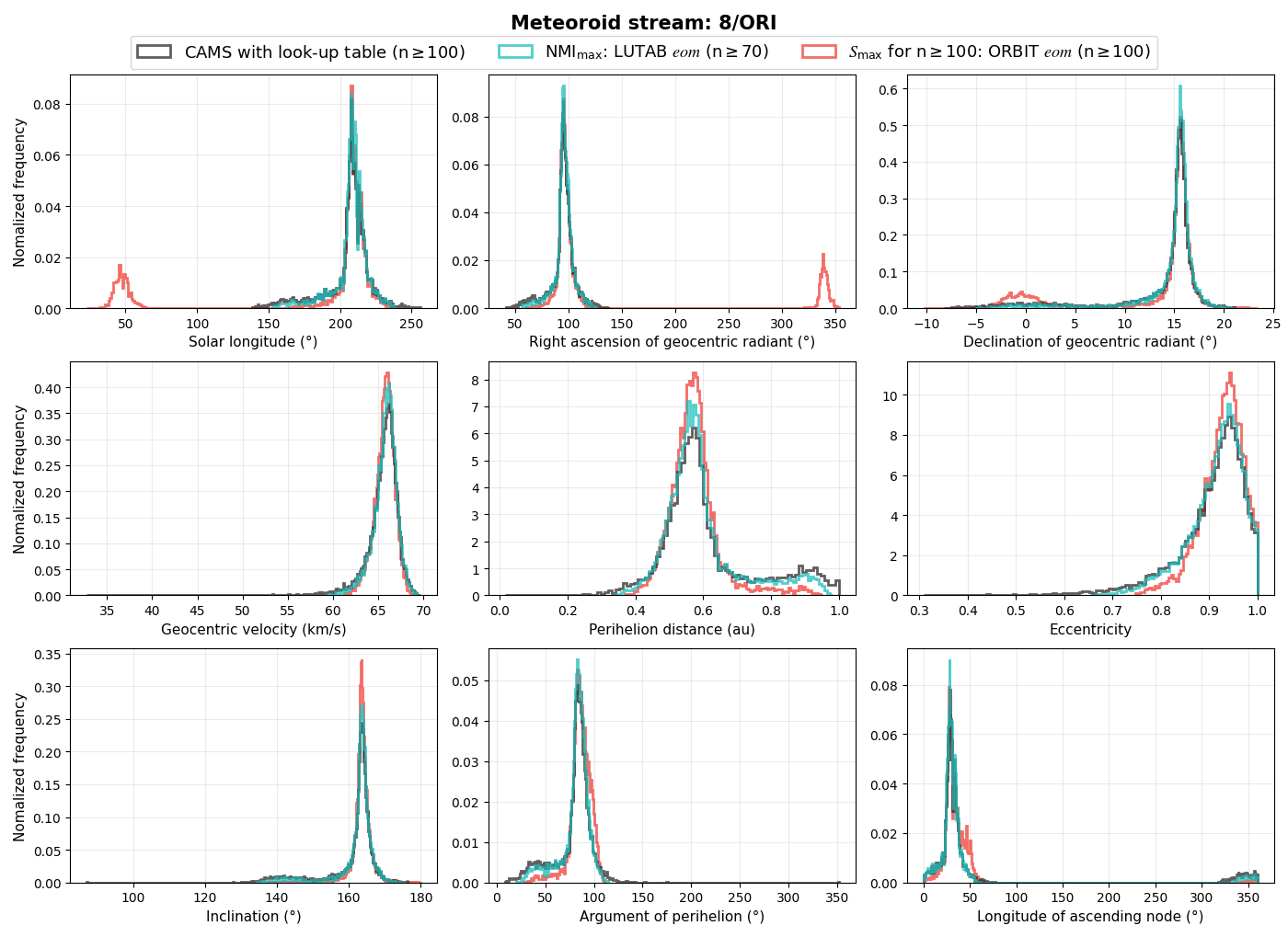}
    \caption{Mean features of the Orionids.}
    \label{fig:shower_ORI}
\end{figure*}

\begin{figure*}[h!]
    \centering
    \includegraphics[width=0.8\textwidth]{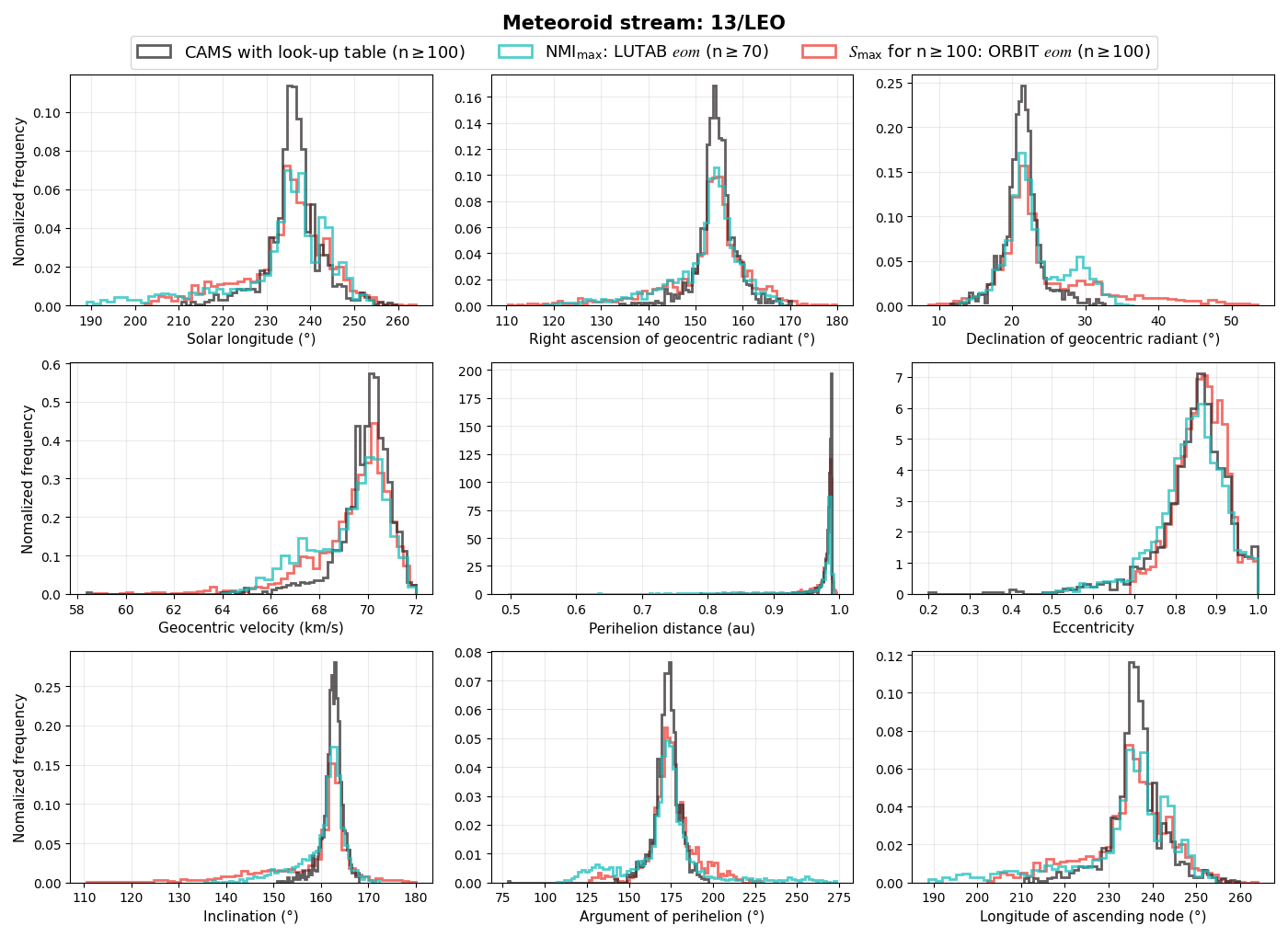}
    \caption{Mean features of the Leonids.}
    \label{fig:shower_LEO}
\end{figure*}

\begin{figure*}[h!]
    \centering
    \includegraphics[width=0.8\textwidth]{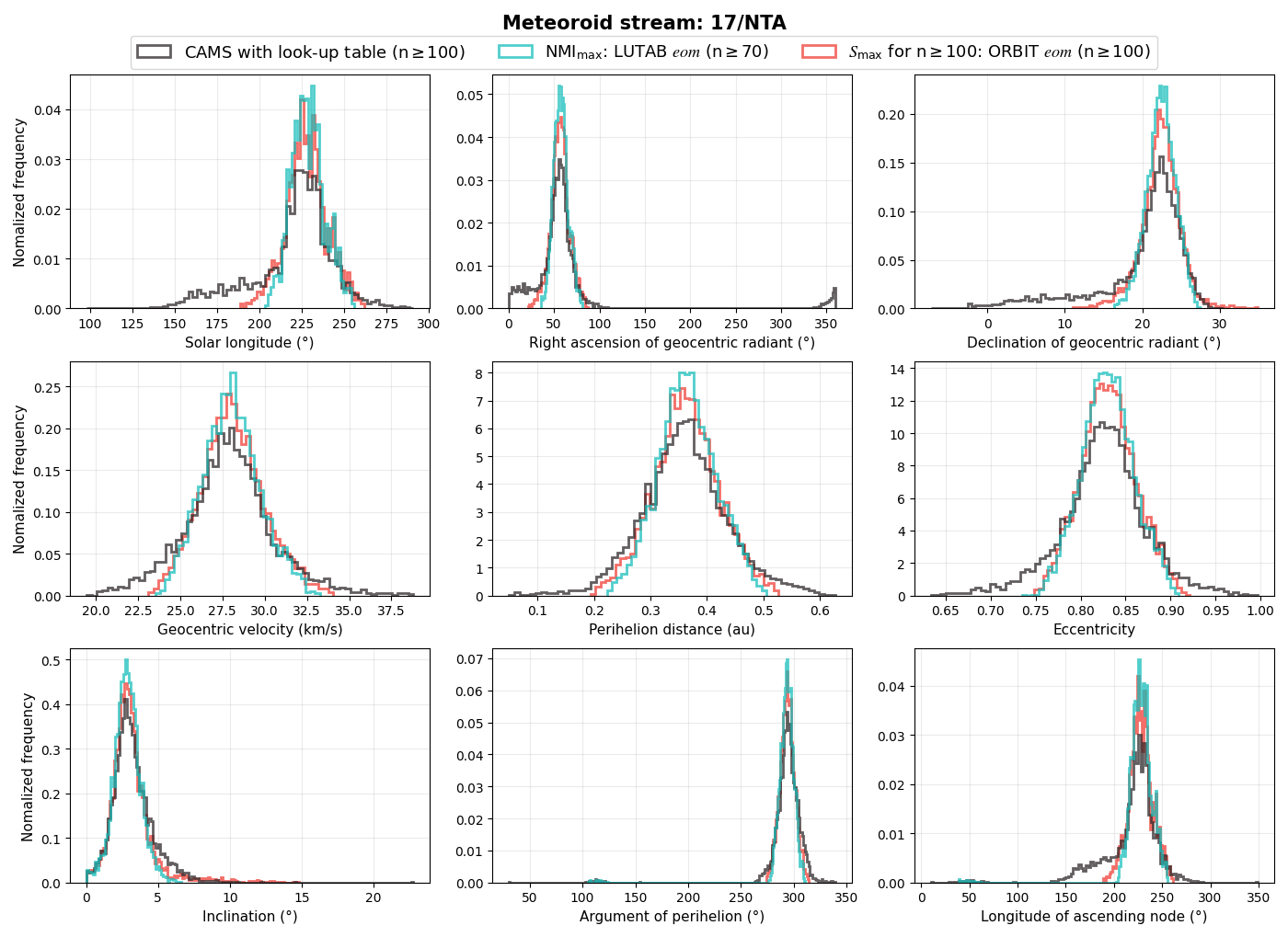}
    \caption{Mean features of the Northern Taurids.}
    \label{fig:shower_NTA}
\end{figure*}

\begin{figure*}[h!]
    \centering
    \includegraphics[width=0.8\textwidth]{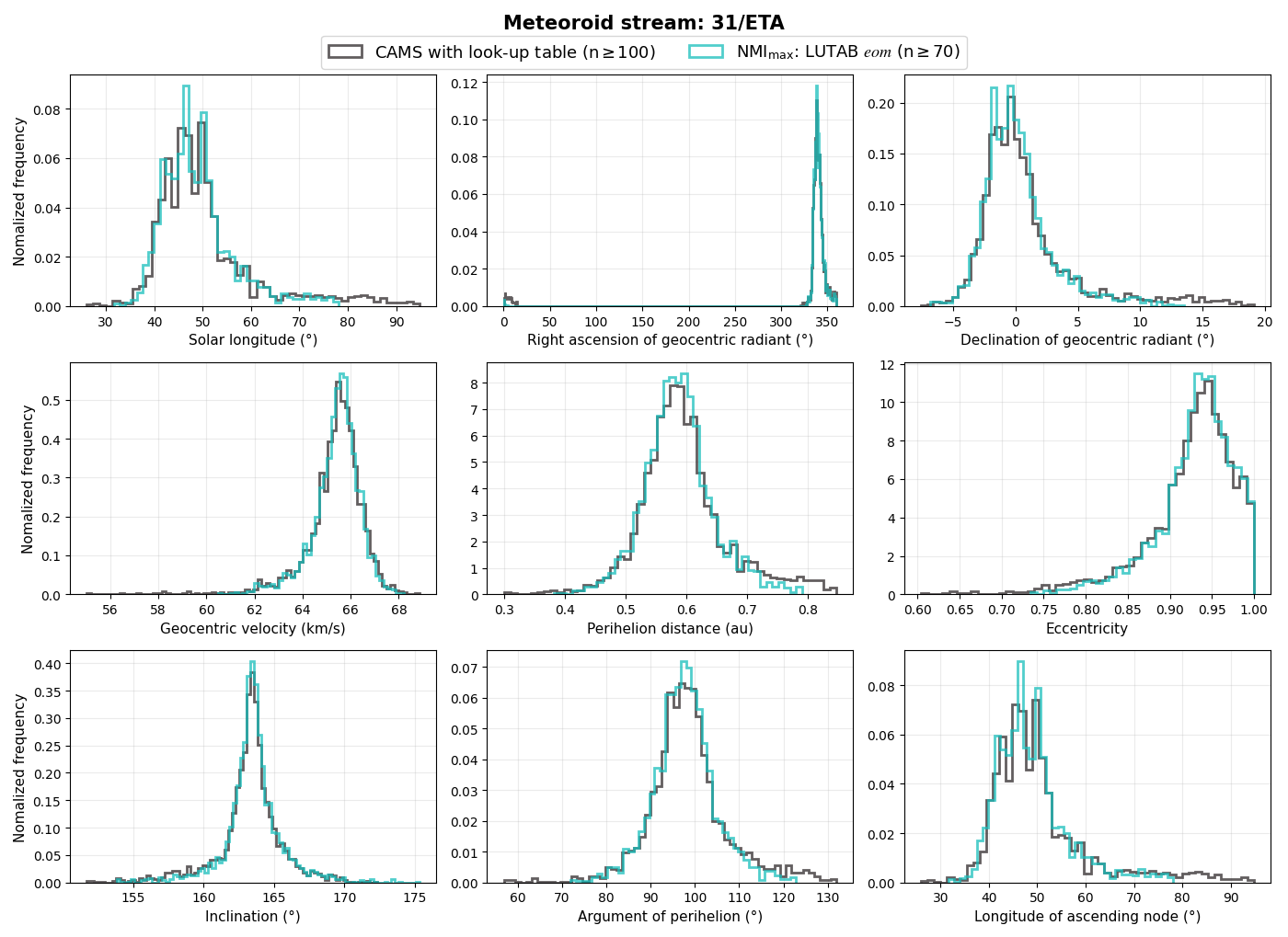}
    \caption{Mean features of the Southern $\eta$-Aquariids.}
    \label{fig:shower_ETA}
\end{figure*}

\begin{figure*}[h!]
    \centering
    \includegraphics[width=0.8\textwidth]{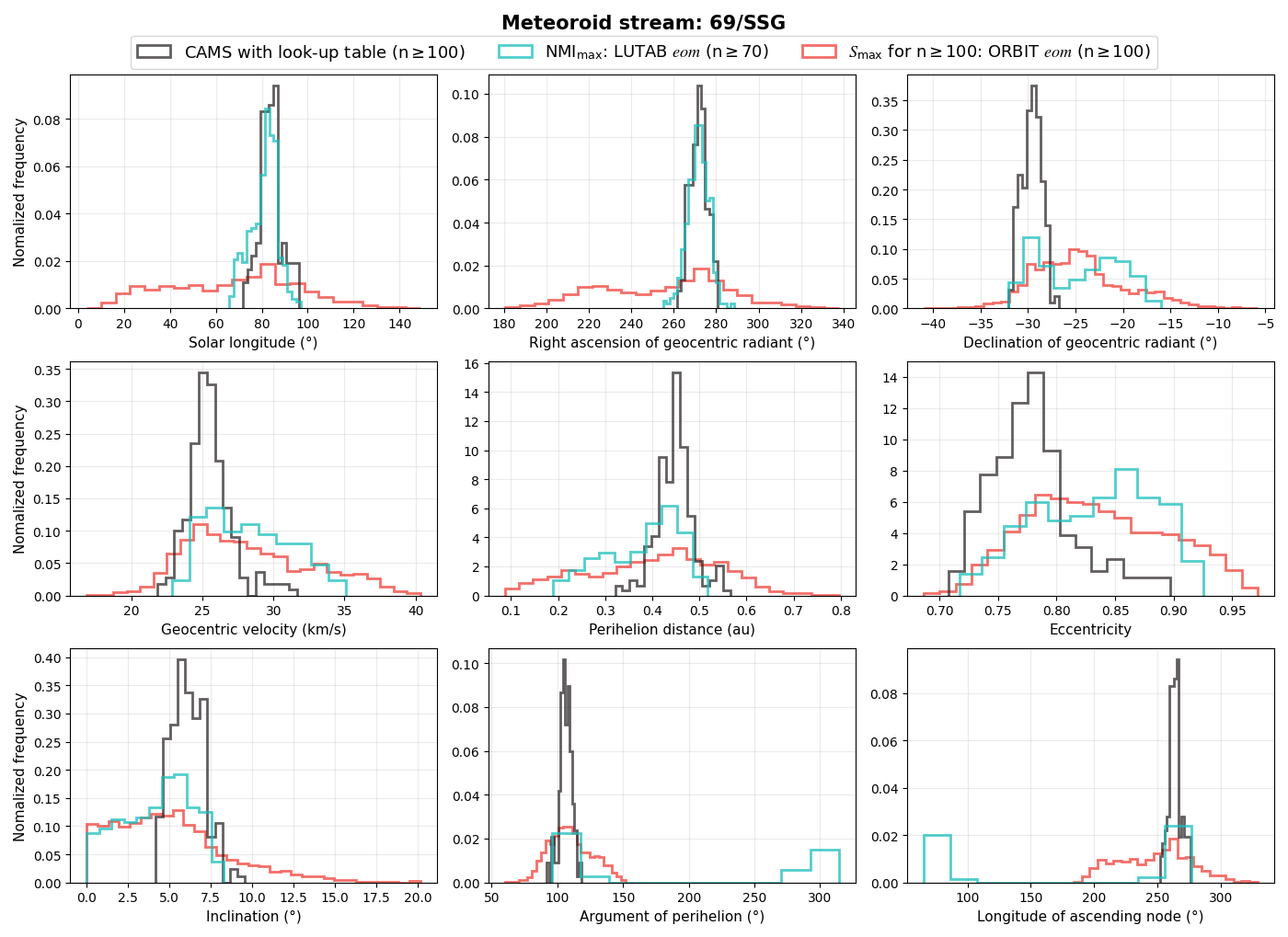}
    \caption{Mean features of the Southern $\mu$-Sagittariids.}
    \label{fig:shower_SSG}
\end{figure*}

\begin{figure*}[h!]
    \centering
    \includegraphics[width=0.8\textwidth]{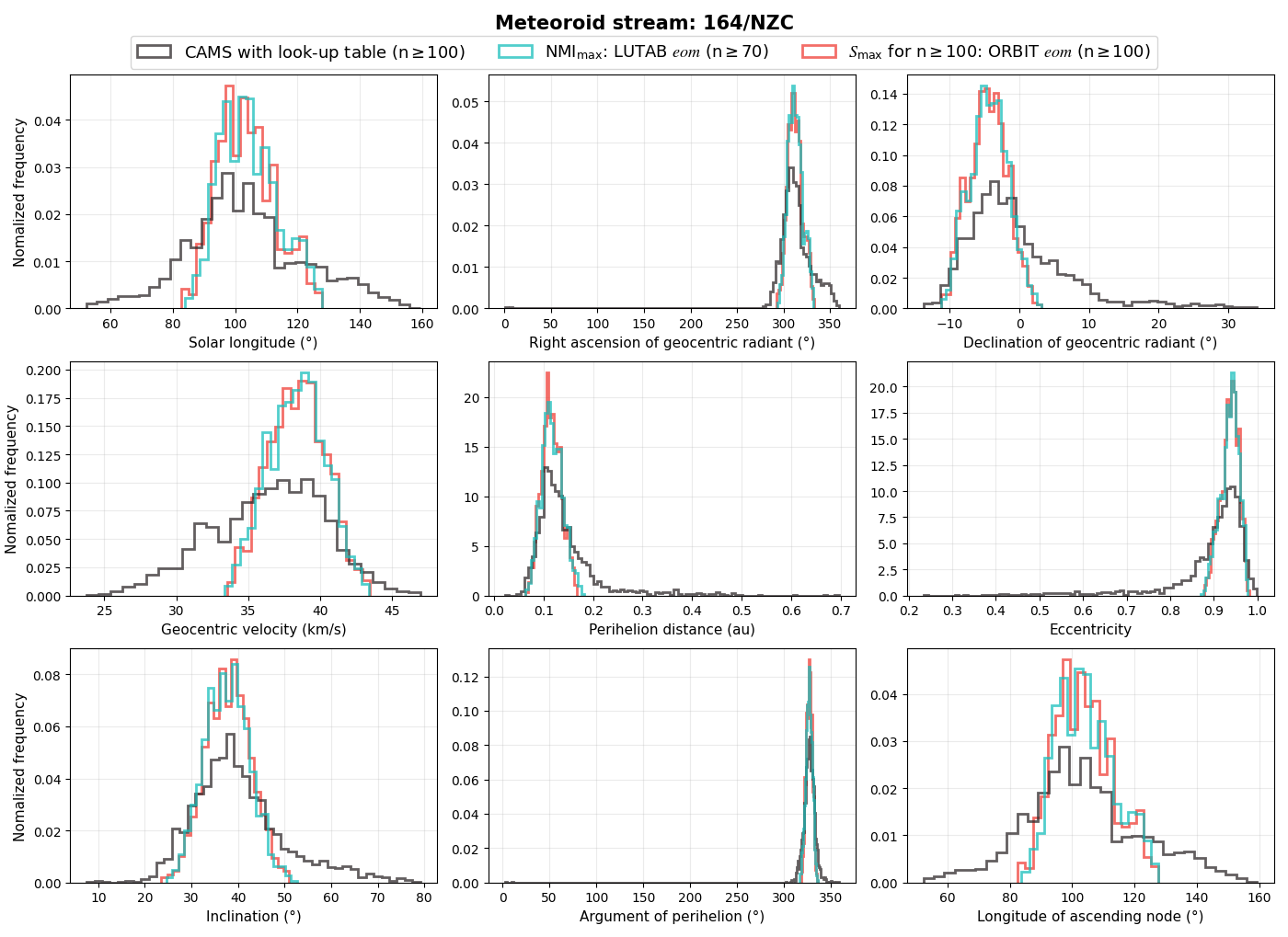}
    \caption{Mean features of the Northern June Aquilids.}
    \label{fig:shower_NZC}
\end{figure*}

\begin{figure*}[h!]
    \centering
    \includegraphics[width=0.8\textwidth]{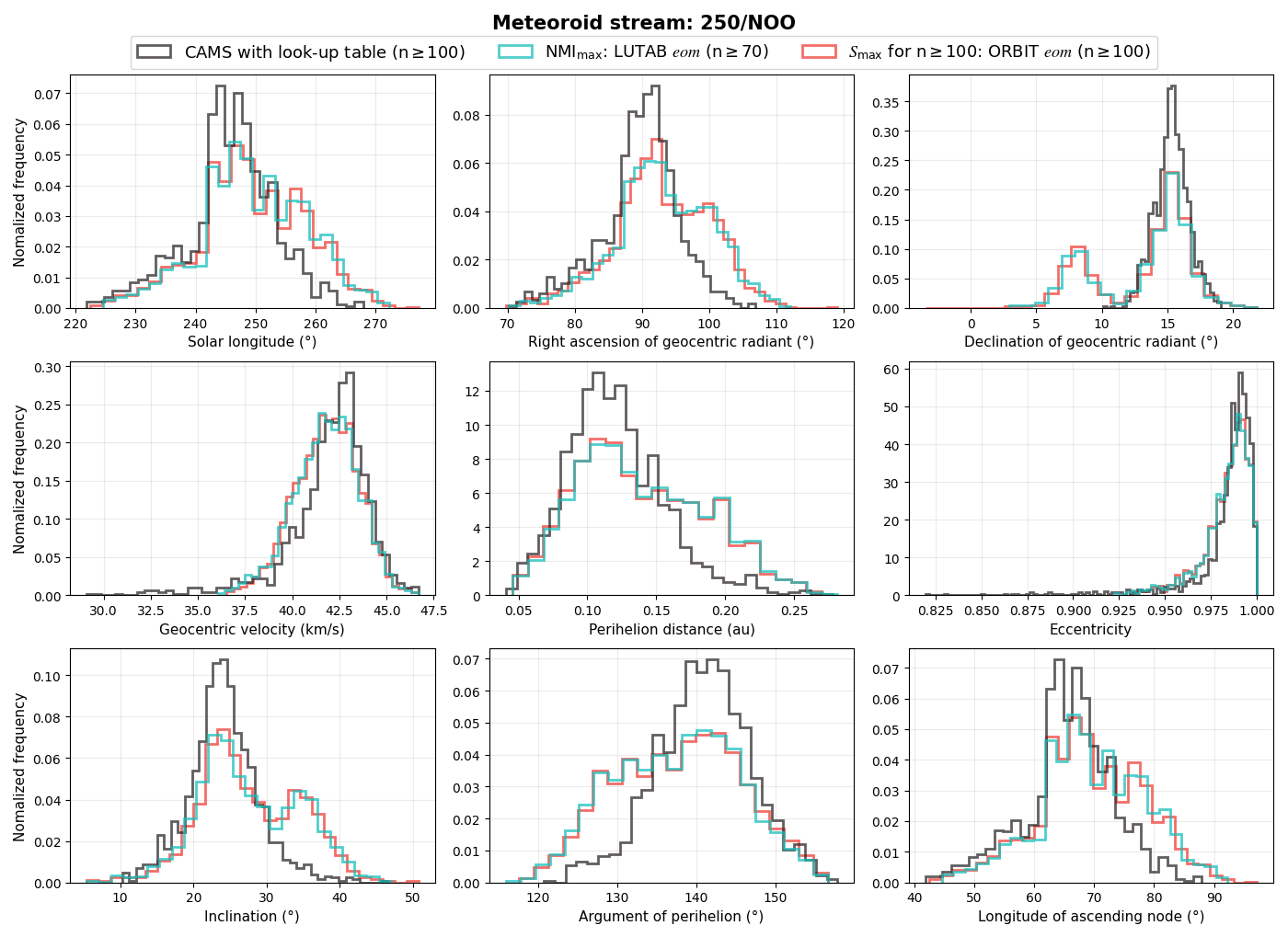}
    \caption{Mean features of the November Orionids.}
    \label{fig:shower_NOO}
\end{figure*}

\begin{figure*}[h!]
    \centering
    \includegraphics[width=0.8\textwidth]{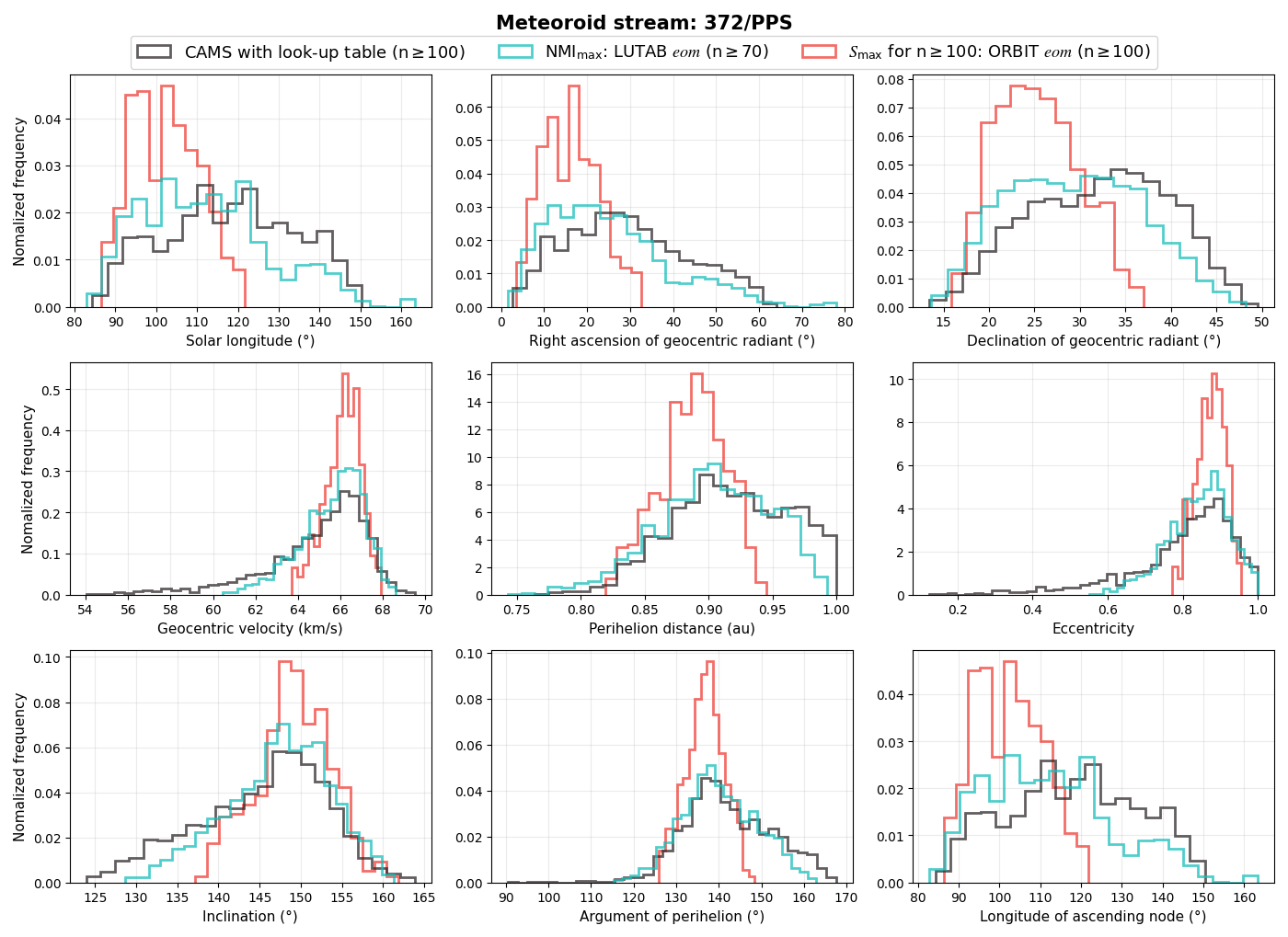}
    \caption{Mean features of the $\phi$-Piscids.}
    \label{fig:shower_PPS}
\end{figure*}

\end{document}